\documentclass[12pt]{article}
\usepackage{epsfig}
\usepackage{amssymb}
\usepackage{pslatex}
\usepackage[usenames]{color}

\makeatletter
\renewcommand{\theequation}{\thesection.\arabic{equation}}
\@addtoreset{equation}{section}
\makeatother
\newcommand{\bea}{\begin{eqnarray}}
\newcommand{\eea}{\end{eqnarray}}
\newcommand{\be}{\begin{eqnarray}}
\newcommand{\ee}{\end{eqnarray}}

\def\tilde{\widetilde}
\def\bar{\overline}
\def\rmd{{\rm d}}

\begin{document}

\begin{titlepage}
\vskip1cm
%\begin{flushright}
%\end{flushright}
%
\centerline{\Large \bf  Exact Results and Holography of Wilson Loops}
\vspace{0.3cm}
\centerline{\Large \bf in}
\vspace{0.3cm}
\centerline{\Large \bf ${\cal N}=2$ Superconformal (Quiver) Gauge Theories}
\vspace{1.25cm}
\centerline{\large Soo-Jong Rey $^{a,b}$, \,\,\, Takao Suyama $^a$}
\vspace{1.25cm}
\centerline{\sl $^a$ School of Physics and Astronomy \& Center for Theoretical Physics}
\vskip0.25cm
\centerline{\sl Seoul National University, Seoul 141-747 {\rm KOREA}}
\vskip0.25cm
\centerline{\sl $^b$ School of Natural Sciences, Institute for Advanced Study, Princeton NJ 08540 {\rm USA}}
\vskip0.25cm
\centerline{\tt sjrey@snu.ac.kr \,\,\, suyama@phya.snu.ac.kr}
\vspace{1.0cm}
\centerline{ABSTRACT}
\vspace{0.5cm}
\noindent
Using localization, matrix model and saddle-point techniques, we determine exact behavior of circular Wilson loop in ${\cal N}=2$ superconformal (quiver) gauge theories in the large number limit of colors.
Focusing at planar and large `t Hooft couling limits, we compare its asymptotic behavior with well-known exponential growth of Wilson loop in ${\cal N}=4$ super Yang-Mills theory with respect to `t Hooft coupling. For theory with gauge group SU$(N)$ coupled to $2N$ fundamental hypermultiplets, we find that Wilson loop exhibits {\sl non-exponential} growth -- at most, it can grow as a power of `t Hooft coupling. For theory with gauge group SU($N)\times {\rm SU}(N)$ and bifundamental hypermultiplets, there are two Wilson loops associated with two gauge groups. We find Wilson loop in untwisted sector grows exponentially large as in ${\cal N}=4$ super Yang-Mills theory. We then find Wilson loop in twisted sector exhibits {\sl non-analytic} behavior with respect to difference of the two `t Hooft coupling constants. By letting one gauge coupling constant hierarchically larger/smaller than the other, we show that Wilson loops in
the second type theory interpolate to Wilson loops in the first type theory.
We infer implications of these findings from holographic dual description in terms of minimal surface of dual string worldsheet. We suggest intuitive interpretation that in both classes of theory holographic dual background must involve string scale geometry even at planar and large `t Hooft coupling limit and that new results found in the gauge theory side are attributable to worldsheet instantons and infinite resummation therein. Our interpretation also indicates that holographic dual of these gauge theories is provided by certain {\sl non-critical} string theories.
\end{titlepage}

\newpage

%%%%%%%%%%%%%%%%%%%%%%%%%%%%%%%%%%%%%%%%%%%%%%%%%%%%%%%%%%%%%%%%%%%%%%%%%%%%%%%%%%%%%%%%%%%%%%%%%%%%%%%%%%%%%
\section{Introduction}

\vspace{5mm}

AdS/CFT correspondence \cite{Maldacena:1997re} between ${\cal N}=4$ super Yang-Mills theory and Type IIB string theory on $AdS_5\times S^5$ has been studied extensively during the last decade. One remarkable result obtained from the study is exact computation for expectation value of Wilson loop operators at strong coupling \cite{Rey:1998ik}\cite{Maldacena:1998im}. For a half-BPS circular Wilson loop, based on perturbative calculations at weak `t Hooft coupling \cite{Erickson:2000af}, exact form of the expectation value was conjectured in \cite{Drukker:2000rr}, precisely reproducing the result expected from the string theory computation \cite{Rey:1998ik}, \cite{Maldacena:1998im} and conformal anomaly therein. Their conjecture was confirmed later in \cite{Pestun:2007rz} using a localization technique.

In this paper, we study aspects of half-BPS circular Wilson loops in ${\cal N}=2$ supersymmetric gauge theories. We focus on a class of ${\cal N}=2$ superconformal gauge theories
--- the $A_1$ (quiver) gauge theory of gauge group SU$(N)$ and $2N$ fundamental hypermultiplets and $\hat{A}_1$ quiver gauge theory of gauge group SU($N)\times$SU$(N)$ and bifundamental hypermultiplets --- and compute the Wilson loop expectation value by adapting the localization technique of \cite{Pestun:2007rz}. We then compare the results with the ${\cal N}=4$ super Yang-Mills theory, which is a special limit of the $\hat{A}_0$ quiver gauge theory of gauge group SU($N$) and an adjoint hypermultiplet. Their quiver diagrams are depicted in Fig. 1.
\vskip1cm
\begin{figure}[ht!]
\centering
\includegraphics[scale=0.68]{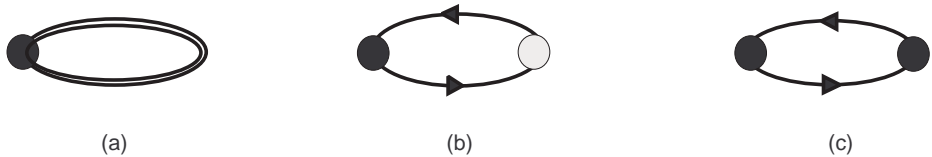}
\caption{\small \sl
Quiver diagram of ${\cal N}=2$ superconformal gauge theories under study: (a) $\hat{A}_0$ theory with $G$ = SU$(N)$ and one adjoint hypermultiplet, (b) $A_1$ theory with $G$=SU($N$) and $2N$ fundamental hypermultiplets, (c) $\hat{A}_1$ theory with $G=$SU($N) \times$ SU($N$) and $2N$ bifundamental hypermultiplets.  The $A_1$ theory is obtainable from $\hat{A}_1$ theory by tuning ratio of coupling constants to 0 or $\infty$. See sections 3 and 4 for explanations.}
\label{}
\end{figure}
\vskip1cm
We show that, on general grounds, path integral of these ${\cal N}=2$ superconformal gauge theories on $\mathbb{S}^4$ is reducible to a finite-dimensional matrix integral. The resulting matrix model turns out very complicated mainly because the one-loop determinant around the localization fixed point is non-trivial. This is in shartp contrast to the ${\cal N}=4$ super Yang-Mills theory, where the one-loop determinant is absent and further evaluation of Wilson loops or correlation functions is straightforward manipulation in Gaussian matrix integral.

Nevertheless, in the $N \rightarrow \infty$ planar limit, we show that expectation value of the half-BPS circular Wilson loop is determinable {\sl provided} the 't Hooft coupling $\lambda$ is large. In the large $\lambda$ limit, the one-loop determinant evaluated by the zeta-function regularization admits a suitable asymptotic expansion. Using this expansion, we can solve the saddle-point equation of the matrix model and obtain large $\lambda$ behavior of the Wilson loop expectation value. In ${\cal N}=4$ super Yang-Mills theory, it is known that the Wilson loop grows exponentially large $ \sim \exp(\sqrt{2 \lambda})$ as $\lambda$ becomes infinitely strong.

In $\hat{A}_0$ gauge theory, we find that the Wilson loop expectation value grows exponentially, exactly the same as the ${\cal N}=4$ super Yang-Mills theory.
The result for $A_1$ gauge theory is surprising. We find that the Wilson loop is finite at large $\lambda$. This means that the Wilson loop exhibits {\sl non-exponential} growth.
The $\hat{A}_1$ quiver gauge theory is also interesting. There are two Wilson loops associated with each gauge groups, equivalently, one in untwisted sector and another in twisted sector. We find that the Wilson loop in untwisted sector scales exponentially large, coinciding with the behavior of the Wilson loop ${\cal N}=4$ super Yang-Mills theory and the $\hat{A}_0$ gauge theory. On the other hand, the Wilson loop in twisted sector exhibits {\sl non-analytic} behavior with respect to difference of two `t Hooft coupling constants. We also find that we can interpolate the two surprising results in $A_1$ and $\hat{A}_1$ gauge theories by tuning the two `t Hooft couplings in $\hat{A}_1$ theory hierarchically different.
In all these, we ignored possible non-perturbative corrections to the Wilson loops. This is because, recalling the fishnet picture for the stringy interpretation of Wilson loops, the perturbative contributions would be the most relevant part for exploring the AdS/CFT correspondence and the holography therein.

We also studied how holographic dual descriptions may explain the exact results. Expectation value of the Wilson loop is described by worldsheet path integral of Type IIB string in dual geometry and that, in case the dual geometry is macroscopically large such as AdS$_5 \times \mathbb{S}^5$, it is evaluated by saddle-points of the path integral -- worldsheet configurations of extremal area surface. We first suggest that non-exponential growth of the $A_1$ Wilson loop arise from delicate cancelation among multiple --- possibly infinitely many --- saddle-points. This implies that holographic dual geometry of the ${\cal N}=2$ $A_1$ gauge theory ought to be (AdS$_5\times {\cal M}_2) \times {\cal M}$ where the internal space ${\cal M}= [\mathbb{S}^1 \times \mathbb{S}^2]$ necessarily involves a geometry of string scale. The string worldsheet sweeps on average an extremal area surface inside AdS$_5$, but many nearby saddle-point configurations whose worldsheet sweep two cycles over ${\cal M}$ cancel among the leading, exponential contributions of each. We next suggest that $\hat{A}_1$ Wilson loop in untwisted sector is given by a macroscopic string in AdS$_5 \times \mathbb{S}^5/\mathbb{Z}_2$ and hence grows exponentially with average of the two `t Hooft coupling constants. In twisted sector, however, it is negligibly small and scales with difference of the two `t Hooft coupling constants. This is again due to delicate cancelation among multiple worldsheet instantons that sweep around collapsed two cycles at the $\mathbb{Z}_2$ orbifold fixed point. We also demonstrate that Wilson loop expectation values are interpolatable between $\hat{A}_1$ and $A_1$ behaviors (or vice versa) by tuning NS-NS 2-form potential on the collapsed two cycle from $1/2$ to $0, 1$ or vice versa.

This paper is organized as follows. In section \ref{sectionN=2}, we show that evaluation of the expectation value of the half-BPS circular Wilson loop in a generic ${\cal N}=2$ superconformal gauge theory reduces to a related problem in a one-matrix model. The reduction procedure is based on localization technique and is parallel to \cite{Pestun:2007rz}. Compared to \cite{Pestun:2007rz}, our derivations are more direct and elementary and hence makes foregoing analysis in the planar limit far clearer physicswise. In section \ref{asymptotic}, we evaluate the Wilson loop at large `t Hooft coupling limit. Based on general analysis for one-matrix model (subsection \ref{generalMM}), we evaluate the matrix model action which is induced by the one-loop determinant (subsection \ref{subseczeta}).
As a result, we obtain a saddle-point equation whose solution provides the large `t Hooft coupling behavior of the Wilson loop (subsection \ref{saddlepoint}).
In section \ref{holography}, we discuss interpretation of these results in holographic dual string theory.
For both $A_1$ and $\hat{A}_1$ types, we argue contribution of worldsheet instanton effects can explain non-analytic behavior of the exact gauge theory results.
Section \ref{discuss} is devoted to discussion, including a possible implication of the present results to our previous work \cite{Rey:2008bh} (see also \cite{Drukker:2008zx}\cite{Chen:2008bp}) on ABJM theory \cite{Aharony:2008ug}.
We relegated several technical points in the appendices. In appendix A, we summarize Killing spinors on $\mathbb{S}^4$. In appendix B, we work out off-shell closure of supersymmetry algebra. In appendix C, we present asymptotic expansion of the Wilson loop. In appendix D, we present detailed computation of $c_1$ that arise in the evaluation of one-loop determinant.

Results of this work were previously reported at KEK workshop and at Strings 2009 conference. For online proceedings, see \cite{suyama} and \cite{rey}, respectively.
\vspace{1cm}

\section{Reduction to One-Matrix Model} \label{sectionN=2}

\vspace{5mm}
The work \cite{Pestun:2007rz} provided a proof for the conjecture \cite{Erickson:2000af, Drukker:2000rr} that the evaluation of the half-BPS Wilson loop in ${\cal N}=4$ super Yang-Mills theory \cite{Rey:1998ik, Maldacena:1998im} is reduced to a related problem in a Gaussian Hermitian one-matrix model. In this section, we show that the similar reduction also works for ${\cal N}=2$ superconformal gauge theories of general quiver type. The resulting matrix model is, however, not Gaussian but includes non-trivial vertices due to {\color{red} nontrivial } one-loop determinant.

\vspace{5mm}

\subsection{From ${\cal N}=4$ to ${\cal N}=2$}

\vspace{5mm}

A shortcut route to an ${\cal N}=2$ gauge theory of general quiver type --- with matters in various different representations and coupling constants in different values --- is to start with ${\cal N}=4$ super Yang-Mills theory. In this section, for completeness of our treatment, we elaborate on this route. Let $G$ be the gauge group.
The latter theory consists of a gauge field $A_m$ with $m=1,2,3,4$, scalar fields $A_0,A_5,\cdots,A_9$ and an $SO(9,1)$ Majorana-Weyl spinor $\Psi$, all in the adjoint representation of $G$.
The action can be written compactly as
\begin{equation}
S_{{\cal N}=4} = \int_{\mathbb{R}^4} \rmd^4 x\ \, \mbox{Tr}\Bigl( -\frac14F_{MN}F^{MN}-\frac i2\bar{\Psi}\Gamma^M D_M\Psi \Bigr),
   \label{N=4}
\end{equation}
where $M,N=0,\cdots,9$ and
\begin{eqnarray}
F_{MN} &=& \partial_MA_N-\partial_NA_M-ig[A_M,A_N], \\
D_M\Psi &=& \partial_M\Psi-ig[A_M,\Psi], \\
\Gamma \Psi &=& +\Psi.
\end{eqnarray}
Note that the metric of the base manifold $\mathbb{R}^4$ is taken in the Euclidean signature, while the ten-dimensional 'metric' $\eta^{MN}$ is taken Lorentzian with $\eta^{00}=-1$.
As usual in the dimensional reduction, the derivatives other than $\partial_m$ are set to zero.

The action (\ref{N=4}) is invariant under the supersymmetry transformations
\begin{eqnarray}
\delta A_M &=& -i\bar{\xi}\Gamma_M\Psi,
   \label{SUSY1} \\
\delta \Psi &=& \frac12F_{MN}\Gamma^{MN}\xi,
   \label{SUSY2}
\end{eqnarray}
where $\xi$ is a constant $SO(9,1)$ Majorana-Weyl spinor-valued supersymmetry parameter satisfying the chirality condition $\Gamma\xi=+\xi$.
In what follows, we rewrite the action (\ref{N=4}) so that the resulting action provides a useful guide to deduce the action of an ${\cal N}=2$ gauge theory with hypermultiplet fields of arbitrary representations.

\vspace{5mm}

We first choose which half of the supercharges of the ${\cal N}=4$ supersymmetry is to be preserved.
This choice corresponds to the choice of embedding the SU(2) R-symmetry of ${\cal N}=2$ theory into the SU(4) R-symmetry of the ${\cal N}=4$ theory. Consider one such embedding defined by the matrix
\begin{equation}
M := \left(
\begin{array}{cc}
x_6+ix_7 & -(x_8-ix_9) \\ x_8+ix_9 & \, x_6-ix_7
\end{array}
\right).
   \label{M}
\end{equation}
Its determinant is
\begin{equation}
\det M = (x_6)^2+(x_7)^2+(x_8)^2+(x_9)^2,
\end{equation}
so it is obvious that any transformation of the form
\begin{equation}
M \to g_LMg_R, \hspace{10mm} g_L\in \mbox{SU}(2)_L, \hspace{5mm} g_R\in \mbox{SU}(2)_R
   \label{SO(4)}
\end{equation}
belongs to the SO(4) transformation acting on $(x_6,\cdots,x_9)\in\mathbb{R}^4$.
Note that this transformation preserves the embedding (\ref{M}).
In the ten-dimensional language, SU(4) R-symmetry of the ${\cal N}=4$ theory is realized as the rotational symmetry SO(6) of $\mathbb{R}^6$.
Therefore, one embedding of SU(2) R-symmetry into SU(4) is chosen by selecting SU$(2)_L$ or SU$(2)_R$.
We choose the latter as the R-symmetry of the ${\cal N}=2$ theories.

There is a U(1) subgroup of SU$(2)_L$ generated by $\sigma^3$.
Let $R(\theta)$ be an element of this U(1).
This is $\theta$-rotation in 67-plane and $(-\theta)$-rotation in 89-plane.
In the following, we require that the supercharges preserved in ${\cal N}=2$ theory should be invariant under the $R(\theta)$. For an infinitesimal $\theta$, $R(\theta)$ acts on the supersymmetry transformation parameter $\xi$ as
\begin{equation}
\delta_\theta\xi = -\frac12\theta(\Gamma^6\Gamma^7-\Gamma^8\Gamma^9)\xi.
\end{equation}
Therefore, $\xi$ should satisfy
\begin{equation}
\Gamma^{6789}\xi = -\xi,
   \label{half}
\end{equation}
selecting eight components out of the original sixteen ones.

The scalar fields $A_s$ with $s=6,7,8,9$ can be combined into the doublet $q^\alpha$ ($\alpha = 1, 2$) of SU$(2)_R$ as
\begin{equation}
q^1 := \frac1{\sqrt{2}}(A_6-iA_7), \qquad q^2 := -\frac1{\sqrt{2}}(A_8+iA_9),
\end{equation}
and their conjugates $q_\alpha = (q^\alpha)^\dag$.
Gamma matrices $\gamma^\alpha,\gamma_\alpha$ are defined similarly in terms of $\Gamma^s$.
They satisfy
\begin{equation}
\{\gamma^\alpha,\gamma_\beta\} = 2\delta^\alpha_\beta, \hspace{5mm}
\{\gamma^\alpha,\gamma^\beta\}=0=\{ \gamma_\alpha,\ \gamma_\beta \ \}.
\end{equation}
Note that, for arbitrary vectors $V^s$ and $W^s$, one has
\begin{equation}
V_sW^s = V_\alpha W^\alpha+V^\alpha W_\alpha.
\end{equation}

The Majorana-Weyl spinor $\Psi$ is split into the chirality eigenstates with respect to
$\Gamma^{6789}$ as follows:
\begin{equation}
\lambda := \frac{1}{2} (1-\Gamma^{6789}) \ \Psi, \qquad \quad \eta := \frac{1}{2} (1+\Gamma^{6789}) \ \Psi.
\end{equation}
Both fermions are Majorana-Weyl. We further split $\eta$ into $\eta_\pm$, which are eigenstates of
\begin{equation}
\gamma := \frac12[\gamma^\alpha, \gamma_\alpha] = \frac i2(\Gamma^6\Gamma^7-\Gamma^8\Gamma^9) \, .
\end{equation}
Note that $\gamma$ is the generator for $R(\theta)$ and hence satisfies
\begin{equation}
\gamma^2 = \frac12(1+\Gamma^{6789}), \qquad [\gamma,\gamma^\alpha] = + \gamma^\alpha, \qquad
[\gamma,\gamma_\alpha] = -\gamma_\alpha.
\end{equation}
Now, $\eta_\pm$ are not Majorana-Weyl.
In fact, they are related {\sl by charge conjugation}
\begin{equation}
(\eta^A_\pm)^{*} = {\cal C} \, \eta^A_\mp,
\end{equation}
where $A$ is the index for the adjoint representation of $G$ and ${\cal C}$ is the complex conjugation matrix.
So, we shall denote $\eta_-$ by $\psi$. Then, modulo a phase factor, $\eta_+$ is $\psi^\dagger$.

In terms of $A_\mu$ $(\mu=0,\cdots,5)$, $q^\alpha, q_\alpha$, $\lambda$ and $\psi$, the action (\ref{N=4}) can be written as
\begin{eqnarray}
S_{{\cal N}=4}
&=& \int_{\mathbb{R}^4} \rmd^4 x\, \mbox{Tr}\Bigl( -\frac14F_{\mu\nu}F^{\mu\nu}-D_\mu q_\alpha D^\mu q^\alpha
    -\frac i2\bar{\lambda}\Gamma^\mu D_\mu\lambda-i\bar{\psi}\Gamma^\mu D_\mu\psi \nonumber \\
& & -g\bar{\lambda}\gamma^\alpha[q_\alpha,\psi]-g\bar{\psi}\gamma_\alpha[q^\alpha,\lambda]
                  -g^2[q_\alpha,q^\beta][q_\beta,q^\alpha]+\frac12g^2[q_\alpha,q^\alpha][q_\beta,q^\beta] \Bigr),
   \label{N=4'}
\end{eqnarray}
with the understanding that the dimensional reduction sets $\partial_\mu = 0$ for $\mu =0,5$.
The supersymmetry transformations (\ref{SUSY1}),(\ref{SUSY2}) can be written as
\begin{eqnarray}
\delta A_\mu &=& -i\bar{\xi}\Gamma_\mu\lambda, \\
\delta q^\alpha &=& -i\bar{\xi}\gamma^\alpha\psi, \\
\delta q_\alpha &=& - i \overline{\psi} \gamma_\alpha \xi \\
\delta \lambda &=& + \frac12F_{\mu\nu}\Gamma^{\mu\nu}\xi-ig[q_\alpha,q^\beta]\gamma^\alpha{}_\beta\xi, \\
\delta \psi &=& + D_\mu q^\alpha\Gamma^\mu\gamma_\alpha\xi.
\end{eqnarray}
Again, if $\xi$ obeys the projection condition (\ref{half}), the action (\ref{N=4'}) has ${\cal N}=2$ supersymmetry.

\vspace{5mm}
At this stage, we shall be explicit of representation contents of $(q^\alpha, \psi)$ fields and their conjugates.
Let $(T^A)^B_C=-if^{AB}_C$ be the generators of Lie$(G)$ in the adjoint representation. We also impose on $\xi$ the projection condition (\ref{half}). In terms of them, the action (\ref{N=4'}) can be written as
\begin{eqnarray}
S_{{\cal N}=2}
&=& \int_{\mathbb{R}^4} \rmd^4 x \, \Bigl( -\frac14\mbox{tr}(F_{\mu\nu}F^{\mu\nu})
    -\frac i2\mbox{tr}(\bar{\lambda}\Gamma^\mu D_\mu\lambda) - D_\mu q_\alpha D^\mu q^\alpha
    -i\bar{\psi}\Gamma^\mu D_\mu\psi \nonumber \\
& & \hspace*{0.3cm} +g\bar{\lambda}^A\gamma^\alpha q_\alpha T_A\psi+g\bar{\psi}\gamma_\alpha T_Aq^\alpha \lambda^A
                  -g^2(q_\alpha T^Aq^\beta)^2+\frac12g^2(q_\alpha T_Aq^\alpha)^2 \Bigr),
   \label{N=2}
\end{eqnarray}
where the gauge covariant derivatives are
\begin{eqnarray}
D_\mu q^\alpha &=& \partial_\mu q^\alpha-iA_\mu^AT_Aq^\alpha, \\
D_\mu q_\alpha &=& \partial_\mu q_\alpha+iq_\alpha T_AA_\mu^A, \\
D_\mu\psi &=& \partial_\mu\psi-iA_\mu^AT_A\psi.
\end{eqnarray}
The ${\cal N}=2$ supersymmetry transformation rules are
\begin{eqnarray}
\delta A_\mu &=& -i\bar{\xi}\Gamma_\mu\lambda,
   \label{SUSY3} \\
\delta \lambda^A &=& + \frac12F^A_{\mu\nu}\Gamma^{\mu\nu}\xi+iq_\alpha T^Aq^\beta\gamma^\alpha{}_\beta\xi, \\
\delta q^\alpha &=& -i\bar{\xi}\gamma^\alpha\psi, \\
\delta q_\alpha &=& -i \bar{\psi} \gamma_\alpha \xi \\
\delta \psi &=& + D_\mu q^\alpha\Gamma^\mu\gamma_\alpha\xi.
   \label{SUSY4}
\end{eqnarray}
The above action (\ref{N=2}) is equivalent to the original action (\ref{N=4}):  we have just rewritten the original action in terms of renamed component fields.
The supersymmetry transformations (\ref{SUSY3})-(\ref{SUSY4}) are also equivalent to
(\ref{SUSY1}) - (\ref{SUSY2}) in so far as $\xi$ is projected to ${\cal N}=2$ supersymmetry as (\ref{half}).

\vspace{5mm}

It turns out that the action (\ref{N=2}) is invariant under ${\cal N}=2$ supersymmetry transformations (\ref{SUSY3})-(\ref{SUSY4}) even for $T^A$ in a generic representation $R$ of the gauge group $G$, which
can also be reducible. Therefore, (\ref{N=2}) defines an ${\cal N}=2$ gauge theory with matter fields $(q^\alpha, \psi)$ in the representation $R$ and their conjugates.

It is also possible to treat $\hat{A}_{k-1}$ quiver gauge theories on the same footing. We embed the orbifold action $\mathbb{Z}_k$ into SU$(2)_L$. In this paper, we shall focus on $\hat{A}_1$ quiver gauge theory. In this case, we should substitute
\begin{eqnarray}
A_\mu = \left(
\begin{array}{cc}
{A_\mu}^{(1)} & \\ & {A_\mu}^{(2)}
\end{array}
\right), &\hspace{5mm}&
\lambda = \left(
\begin{array}{cc}
\lambda^{(1)} \, & \\ & \, \lambda^{(2)}
\end{array}
\right), \nonumber \\
q^\alpha = \left(
\begin{array}{cc}
 & q^{(1)\alpha} \\ q^{(2)\alpha} &
\end{array}
\right), &\hspace{5mm}&
\psi = \left(
\begin{array}{cc}
 & \psi^{(1)} \\ \psi^{(2)} &
\end{array}
\right).
\end{eqnarray}
into (\ref{N=4'}).
Note that the ${\cal N}=2$ supersymmetry (\ref{SUSY3})-(\ref{SUSY4}) is preserved even when the gauge coupling constant $g$ is replaced with
the matrix-valued one:
\begin{equation}
g = \left(
\begin{array}{cc}
g_1 \, \mathbb{I} & \\ & g_2 \, \mathbb{I}
\end{array}
\right).
\end{equation}
In general, $g_1 \ne g_2$ and can be extended to complex domain. Extension to $\hat{A}_k (k \ge 2)$ is straightforward.

\vspace{1cm}

\subsection{Superconformal symmetry on $\mathbb{S}^4$}

\vspace{5mm}

Following \cite{Pestun:2007rz}, we now define the ${\cal N}=2$ superconformal gauge theory on $\mathbb{S}^4$ of radius $r$. For definiteness, we consider the round-sphere with the metric $h_{mn}$ induced through the standard stereographic projection. Details are summarized in Appendix \ref{spinor}.

For this purpose, it also turns out convenient to start with ${\cal N}=4$ super Yang-Mills theory defined on $\mathbb{S}^4$. To maintain conformal invariance, the scalars ought to have the conformal coupling to the curvature scalar of $\mathbb{S}^4$. The action thus reads
\begin{equation}
S_{{\cal N}=4} = \int_{\mathbb{S}^4}  \rmd^4 x \, \sqrt{h}\ \, \mbox{Tr}\Bigl( -\frac14F_{MN}F^{MN}-\frac1{r^2}A_SA^S -\frac i2\bar{\Psi}\Gamma^M D_M\Psi \Bigr),
   \label{N=4SC}
\end{equation}
where $S=0,5,6,\cdots,9$.
The action is invariant under the ${\cal N}=4$ supersymmetry transformations
\begin{eqnarray}
\delta A_M \! &=& -i \, \overline{\xi}\Gamma_M\Psi, \\
\delta \Psi \, &=& + \frac12F_{MN}\Gamma^{MN}\xi-2\Gamma^SA_S\widetilde{\xi} \ ,
\end{eqnarray}
provided that $\xi$ and $\widetilde{\xi}$ satisfy the conformal Killing equations:
\begin{equation}
\nabla_m\xi = \Gamma_m\widetilde{\xi}, \qquad \nabla_m\widetilde{\xi} = -\frac1{4r^2}\Gamma_m\xi.
   \label{Killing}
\end{equation}
Explicit form of the solution to these equations are given in Appendix \ref{spinor}.

The action of an ${\cal N}=2$ gauge theory on $\mathbb{S}^4$ with a hypermultiplet of representation $R$
can be deduced easily as in the previous subsection. One obtains
\begin{eqnarray}
S_{{\cal N}=2}
&=& \int_{{\mathbb{ S}}^4} \rmd^4 x\, \sqrt{h} \, \Bigl( -\frac14\mbox{Tr}(F_{\mu\nu}F^{\mu\nu})
    -\frac i2\mbox{Tr}(\bar{\lambda}\Gamma^\mu D_\mu\lambda) -\frac1{r^2}\mbox{Tr}(A_aA^a)
    \nonumber \\
& & \hspace*{1cm} - \, D_\mu q_\alpha D^\mu q^\alpha -i\bar{\psi}\Gamma^\mu D_\mu\psi -\frac2{r^2}q_\alpha q^\alpha \nonumber \\
& & \hspace*{1cm} + \, g\bar{\lambda}^A\gamma^\alpha q_\alpha T_A\psi+g\bar{\psi}\gamma_\alpha T_Aq^\alpha \lambda^A
                  -g^2(q_\alpha T^Aq^\beta)^2+\frac12g^2(q_\alpha T_Aq^\alpha)^2 \Bigr),
\label{action-on-s4}
\end{eqnarray}
where $a=0,5$.
The action is invariant under the ${\cal N}=2$ superconformal symmetry
\begin{eqnarray}
\delta A_\mu &=& -i\, \overline{\xi}\Gamma_\mu\lambda, \nonumber \\
\delta \lambda^A &=&  + \frac12F^A_{\mu\nu}\Gamma^{\mu\nu}\xi+igq_\alpha T^Aq^\beta\gamma^\alpha{}_\beta\xi
                     -2\Gamma^aA_a^A\widetilde{\xi}, \nonumber \\
\delta q^\alpha &=& -i\, \overline{\xi}\gamma^\alpha \psi, \nonumber \\
\delta q_\alpha &=& - i \, \overline{\psi} \gamma_\alpha \xi \nonumber \\
\delta \,\psi \, &=& + D_\mu q^\alpha\Gamma^\mu\gamma_\alpha\xi-2\gamma_\alpha q^\alpha\widetilde{\xi} \, ,
\nonumber
\end{eqnarray}
where $\xi$ satisfies the conformal Killing equations (\ref{Killing}) in addition to the projection condition (\ref{half}). We emphasize that this is the transformation of the ${\cal N}=2$ superconformal symmetry, not just the Poincar\'e part of it. This can be checked explicitly, for example, by examining the commutator of two transformations on the fields.

\vspace{5mm}

We find it convenient to define a fermionic transformation $Q$ corresponding to the above superconformal transformation $\delta$. It is obtained easily by the replacement $\delta\to\theta Q$ and $\xi\to\theta\xi$ with $\theta$ a real Grassmann parameter.
The resulting transformation is
\begin{eqnarray}
Q A_\mu &=& -i\overline{\xi}\Gamma_\mu\lambda, \nonumber \\
Q \lambda^A &=& + \frac12F^A_{\mu\nu}\Gamma^{\mu\nu}\xi+igq_\alpha T^Aq^\beta\gamma^\alpha{}_\beta\xi
                     -2\Gamma^aA_a^A\widetilde{\xi}, \nonumber \\
Q q^\alpha &=& -i\overline{\xi}\gamma^\alpha\psi, \nonumber \\
Q q_\alpha &=& -i\overline{\psi}\gamma^\alpha\xi, \nonumber \\
Q \psi &=& + D_\mu q^\alpha\Gamma^\mu\gamma_\alpha\xi-2\gamma_\alpha q^\alpha\tilde{\xi},
\end{eqnarray}
where now $\xi$ and $\widetilde{\xi}$ are {\it bosonic} SO(9,1) Majorana-Weyl spinors satisfying ${\cal N}=2$ projection (\ref{half}) and conformal Killing equation (\ref{Killing}).

\vspace{1cm}

\subsection{Localization}

\vspace{5mm}

By extending the localization technique of \cite{Pestun:2007rz}, we now show that computation of Wilson loop expectation value in ${\cal N}=2$ superconformal gauge theory of quiver type can be reduced to computation of a one-matrix integral.

Let $\mathfrak{Q}$ be a fermionic transformation. Suppose that an action $S$ under consideration is invariant under $\mathfrak{Q}$. Then, the following modification
\begin{equation}
S(t) := S + t\int \rmd^4x \, \sqrt{h}\, \mathfrak{Q} V(x)
\end{equation}
does not change the partition function provided that
\begin{equation}
\int \rmd^4x\, \sqrt{h}\, \mathfrak{Q}^2 V(x) = 0.
   \label{Q^2V}
\end{equation}
Likewise, correlation functions remain unchanged if operators under consideration are $\mathfrak{Q}$-invariant.
We shall choose $V(x)$ such that the bosonic part of $\mathfrak{Q} V(x)$ is positive semi-definite.
For this choice, since $t$ can be chosen to be an arbitrary value, we can take the limit $t\to+\infty$ so that the path-integral is localized to configurations where the bosonic part of $\mathfrak{Q} V(x)$ vanishes.
It will turn out later that the vanishing locus of $\mathfrak{Q} V(x)$ is parametrized by a constant matrix. This is why the evaluation of the expectation value of a $\mathfrak{Q}$-invariant operator reduces to a matrix integral. The action of the resulting
matrix model is the sum of $S$ evaluated at the vanishing locus and the one-loop determinant obtained from the quadratic terms of $\mathfrak{Q} V(x)$ when expanded around the vanishing locus.

\vspace{5mm}

One might think that the fermionic transformation $Q$ defined in the previous section can be used as $\mathfrak{Q}$ above. In fact, $Q^2$ is a sum of bosonic transformations, and therefore, (\ref{Q^2V}) appears to hold as long as $V(x)$ is invariant under the transformations. The problem of this choice is that $Q^2$ is such a sum only on-shell. According to \cite{Berkovits:1993zz},\cite{Evans:1994cb} and \cite{Baulieu:2007ew}, $Q$ has to be modified so that the resulting $\mathfrak{Q}$ closes to a sum of bosonic transformations for off-shell.

To this end, we introduce auxiliary fields $K^{\dot{m}}$ $(\dot{m}=\hat{2},\hat{3},\hat{4})$, $K^\alpha$ and $K_\alpha$. They transform in the adjoint, $R$ and $\bar{R}$ representations of the gauge group $G$, respectively.
Utilizing them, we modify the action (\ref{action-on-s4}) in a trivial manner:
\begin{eqnarray}
S_{{\cal N}=2}
&=& \int_{\mathbb{S}^4} \rmd^4 x\, \Bigl( -\frac14\mbox{Tr}(F_{\mu\nu}F^{\mu\nu})
    -\frac i2\mbox{Tr}(\bar{\lambda}\Gamma^\mu D_\mu\lambda)-\frac1{r^2}\mbox{Tr}(A_aA^a) \nonumber \\
    && \hspace*{1.5cm} -\, D_\mu q_\alpha D^\mu q^\alpha
    -i\bar{\psi}\Gamma^\mu D_\mu\psi -\frac2{r^2}q_\alpha q^\alpha \nonumber \\
& & \hspace*{1.5cm} +\, g\bar{\lambda}^A\gamma^\alpha q_\alpha T_A\psi+g\bar{\psi}\gamma_\alpha T_Aq^\alpha \lambda^A -g^2(q_\alpha T^Aq^\beta)^2+\frac12g^2(q_\alpha T_Aq^\alpha)^2 \nonumber \\
& & \hspace*{1.5cm} +\, \frac12K^{\dot{m}}K_{\dot{m}}+K_\alpha K^\alpha \Bigr).
   \label{modified}
\end{eqnarray}
Evidently, this action is physically equivalent to the original one.
The modified action (\ref{modified}) is now invariant under the following $\mathfrak{Q}$ transformations:
\begin{eqnarray}
\mathfrak{Q} \, A_\mu \, &=& -i\overline{\xi}\Gamma_\mu\lambda, \nonumber \\
\mathfrak{Q} \, \lambda^A &=& + \frac12F^A_{\mu\nu}\Gamma^{\mu\nu}\xi+igq_\alpha T^Aq^\beta\gamma^\alpha{}_\beta\xi
                     -2\Gamma^aA_a^A\widetilde{\xi}+K^{\dot{m}A}\nu_{\dot{m}}, \nonumber \\
\mathfrak{Q} \,  q^\alpha \, &=& -i\overline{\xi}\gamma^\alpha\psi, \nonumber \\
\mathfrak{Q} \, q_\alpha &=& -i\overline{\psi}\gamma_\alpha\xi, \nonumber \\
\mathfrak{Q} \, \psi \,\, &=& + D_\mu q^\alpha\Gamma^\mu\gamma_\alpha\xi-2\gamma_\alpha q^\alpha\widetilde{\xi}+K^\alpha\nu_\alpha, \nonumber \\
\mathfrak{Q} \, \overline{\psi} \,\, &=& + D_\mu q_\alpha\bar{\xi}\gamma^\alpha\Gamma^\mu+2\bar{\widetilde{\xi}}\gamma^\alpha q_\alpha
                 +K_\alpha\overline{\nu}^\alpha, \nonumber \\
\mathfrak{Q} K^{\dot{m}A} \!\!\! &=& -\overline{\nu}^{\dot{m}}\Bigl( -i\Gamma^\mu D_\mu\lambda^A+g\gamma^\alpha q_\alpha T^A\psi
                 -g\gamma_\alpha\psi^*T^Aq^\alpha \Bigr), \nonumber \\
\mathfrak{Q}  K^\alpha \, &=& -\overline{\nu}^\alpha\Bigl( -i\Gamma^\mu D_\mu\psi+\gamma_\beta T_Aq^\beta g\lambda^A \Bigr), \nonumber \\
\mathfrak{Q}  K_\alpha \, &=& -\Bigl( -iD_\mu\overline{\psi}\Gamma^\mu-g\bar{\lambda}^A\gamma^\beta q_\beta T_A \Bigr) \nu_\alpha.
\end{eqnarray}
To make $\mathfrak{Q}^2$ close to a sum of bosonic transformations off-shell, the spinors $\nu^{\dot{m}}$, $\nu^\alpha$, $\overline{\nu}_\alpha$ should be chosen appropriately out of $\xi, \widetilde{\xi}$. Details on them are summarized in Appendix \ref{spinor2}.
With the correct choice, $\mathfrak{Q}^2$ closes, for example, on $\lambda$ as follows:
\begin{equation}
-i\mathfrak{Q}^2\lambda \,
= \, \left( v^m\nabla_m\lambda-\frac12(\overline{\xi}\Gamma_{mn}\widetilde{\xi})\Gamma^{mn}\lambda-ig[v^\mu A_\mu,\lambda] \right)
    +\frac12(\overline{\xi}\Gamma_{st}\widetilde{\xi})\Gamma^{st}\lambda.
\end{equation}
This shows that $\mathfrak{Q}^2$ is a sum of a diffeomorphism on $\mathbb{S}^4$, a $G$ gauge transformation and a global SU$(2)_R$ transformation. In particular, notice that $\overline{\xi}\Gamma_{st}\tilde{\xi}$ turns out to be independent of $x^m$. The action of $\mathfrak{Q}^2$ on the auxiliary fields is slightly different.
For example, on $K^{\dot{m}}$, one obtains
\begin{equation}
-i\mathfrak{Q}^2K^{\dot{m}}
= v^k\nabla_kK^{\dot{m}}-ig[v^\mu A_\mu,K^{\dot{m}}]+\bar{\nu}^{\dot{m}}\Gamma^k\nabla_k\nu_{\dot{n}}K^{\dot{n}}.
\end{equation}
Here, the index $\dot{m}$ does not transform as a part of the four-vector on $\mathbb{S}^4$.
This is not a problem since $K^{\dot{m}}$ is contracted with $\nu_{\dot{m}}$ in $V$ defined below,
and not with some other four-vectors.
The $\mathfrak{Q}$ defined above is the right transformation available for the localization procedure.

\vspace{5mm}
We are at the position to choose $V$. We take
\begin{equation}
V := \mbox{Tr}(V_\lambda\lambda)+V_\psi\psi+\bar{\psi}V_{\bar{\psi}},
\end{equation}
where
\begin{eqnarray}
V_\lambda &=& \frac12F_{\mu\nu}\overline{\xi}\Gamma^0\Gamma^{\mu\nu}+igq_\alpha T^Aq^\beta t_A\overline{\xi}\Gamma^0\gamma^\alpha{}_\beta
 +2\overline{\tilde{\xi}}\Gamma^0\Gamma^aA_a+K^{\dot{m}}\overline{\nu}_{\dot{m}}\Gamma^0, \\
V_\psi &=& D_\mu q_\alpha\overline{\xi}\Gamma^0\Gamma^\mu\gamma^\alpha+2\overline{\tilde{\xi}}\Gamma^0\gamma^\alpha q_\alpha
 +K_\alpha\overline{\nu}^\alpha\Gamma^0, \\
V_{\bar{\psi}} &=& D_\mu q^\alpha\gamma_\alpha\Gamma^\mu\Gamma^0\xi-2\gamma_\alpha q^\alpha\Gamma^0\tilde{\xi}
 +K^\alpha\Gamma^0\nu_\alpha.
\end{eqnarray}
Note that $V$ is a scalar with respect to a particular combination of the diffeomorphism on $\mathbb{S}^4$, the $G$ gauge transformation and the global SU$(2)_R$ transformation. This follows from the identities for the spinors, for example,
\begin{equation}
v^m\nabla_m\xi-\frac12(\overline{\xi}\Gamma_{mn}\widetilde{\xi})\Gamma^{mn}\xi+\frac12(\overline{\xi}\Gamma_{st}
\widetilde{\xi})\Gamma^{st}\xi
= 0,
\end{equation}
and similar ones for $\widetilde{\xi}$ and $\nu^I$ which are summarized in Appendix \ref{spinor} and \ref{spinor2}. Therefore, (\ref{Q^2V}) is satisfied with this choice, as required.

After straightforward but tedious algebra, one obtains the bosonic part of $\mathfrak{Q} V$ expressed as
\begin{eqnarray}
& & \mbox{Tr}(V_\lambda \mathfrak{Q} \lambda)+V_\psi \mathfrak{Q} \psi+\mathfrak{Q} \overline{\psi}V_{\overline{\psi}}\, \Big|_{\rm bosonic} \nonumber \\
&=& \mbox{Tr}\Bigl[ \cos^2\frac\theta2(F^+_{mn}+w^+_{mn}A_5)^2+\sin^2\frac\theta2(F^-_{mn}+w^-_{mn}A_5)^2
    -(K^{\dot{m}}-2A_0\overline{\nu}^{\dot{m}}\tilde{\xi})^2 \nonumber \\
& & +D_mA_aD^mA^a-\frac12g^2[A_a,A_b]^2
    +g^2t_At_B(2q_\alpha T^Aq^\beta q_\beta T^Bq^\alpha-q_\alpha T^A q^\alpha q_\beta T^B q^\beta)
    \Bigr] \nonumber \\
& & +2D_0q_\alpha D^0q^\alpha+2|D_{\dot{\mu}}q^\alpha+\overline{\xi}\Gamma^0{}_{\dot{\mu}}\gamma^\alpha{}_\beta\widetilde{\xi} q^\beta|^2 +\frac3{2r^2}q_\alpha q^\alpha-2K_\alpha K^\alpha,
\end{eqnarray}
where $\theta$ is the polar angle on $\mathbb{S}^4$, $\dot{\mu}=1,2,\cdots,5$ and
\begin{eqnarray}
w^{+}_{mn}
&:=& \frac1{\cos^2\frac\theta2}\overline{\xi}\Gamma^{05}\Gamma_{mn}
\frac{1-\Gamma^{\hat{1}\hat{2}\hat{3}\hat{4}}}2\widetilde{\xi}, \\
w^{-}_{mn}
&:=& \frac1{\sin^2\frac\theta2}\overline{\xi}\Gamma^{05}\Gamma_{mn}
\frac{1+\Gamma^{\hat{1}\hat{2}\hat{3}\hat{4}}}2\widetilde{\xi} \ .
\end{eqnarray}
Here, the hatted indices are the Lorentz ones.
The above expression shows that, after a suitable Wick rotation for $A_0$ and the auxiliary fields, the bosonic part of $\mathfrak{Q} V$ is positive semi-definite.
Therefore, by taking the limit $t\to+\infty$, the path-integral is localized at the vanishing locus of $\mathfrak{Q} V$.
It turns out that, as in \cite{Pestun:2007rz}, non-zero fields at the vanishing locus are
\begin{equation}
A_0 = -\frac i{gr}\Phi, \qquad K^{\hat{2}} = -\frac i{gr^2}\Phi,
   \label{locus}
\end{equation}
where $\Phi$ is a {\sl constant} Hermitian matrix.
The coefficients are chosen for later convenience.

Now, the path-integral is reduced to an integral over the Hermitian matrix $\Phi$.
The action of the corresponding matrix model is a sum of the action (\ref{modified})
evaluated at the vanishing locus and the one-loop determinant for the quadratic terms in $\mathfrak{Q} V$.
Note that higher-loop contributions vanish in the large $t$ limit since $t^{-1}$ plays the role of the loop-counting parameter. At the vanishing locus, the action (\ref{modified}) takes the value
\begin{equation}
S = -\int_{\mathbb{S}^4} \rmd^4 x\, \sqrt{h}\,\mbox{Tr}\Bigl( \frac1{r^2}(A_0)^2+\frac12(K^{\hat{2}})^2 \Bigr) = \frac{4\pi^2}{g^2}\, \mbox{Tr}\, \Phi^2.
\end{equation}
An important difference from the ${\cal N}=4$ super Yang-Mills theory is that the one-loop determinant around the vanishing locus does not cancel and has a complicated functional structure.
In the next section, we show that the presence of the non-trivial one-loop determinant is crucial for determining the large `t Hooft coupling behavior of the half-BPS Wilson loop.

\vspace{5mm}

The half-BPS Wilson loop of ${\cal N}=2$ gauge theory has the following form:
\begin{equation}
W[C] := \mbox{Tr}\, P_s \exp\biggl[ ig\int_0^{2\pi} \rmd s \Bigl( \dot{x}^mA_m(x)+\theta^aA_a(x) \Bigr) \biggr].
\end{equation}
The functions $x^m(s)$, $\theta^a(s)$ are chosen appropriately to preserve a half of the ${\cal N}=2$ superconformal symmetry. We shall choose $C$ to be the great circle at the equator of $\mathbb{S}^4$ (i.e. $\theta=\frac\pi2$) specified by
\begin{equation}
(x^1,x^2,x^3,x^4) = (2r\cos s, 2r\sin s,0,0),
\end{equation}
and $\theta^a$ as
\begin{equation}
\theta^0 = r, \hspace{5mm} \theta^5 = 0.
\end{equation}
For this choice, one can show that
\begin{equation}
\dot{x}^mA_m(x)+\theta^aA_a(x) = -rv^\mu A_\mu(x),
\end{equation}
where $v^\mu=\overline{\xi}\Gamma^\mu\xi$.
See Appendix \ref{spinor} for the explicit expressions of $v^\mu$.
This implies that $W[C]$ is invariant under $\mathfrak{Q}$ due to the identity
\begin{equation}
\overline{\xi}\Gamma^\mu\xi\,\overline{\xi}\Gamma_\mu\lambda = 0.
\end{equation}
Thus, we have shown that $\langle W[C] \rangle$ is calculable by a finite-dimensional matrix integral. The operator whose expectation value in the matrix model is equal to $\langle W[C] \rangle$ is
\begin{equation}
\mbox{Tr}\, \exp\Bigl( 2\pi \Phi \Bigr).
\end{equation}
Notice that it is solely governed by the constant-valued, Hermitian matrix $\Phi$. This enables us to compute the Wilson loops in terms of a matrix integral. This observation will also play a role in identifying holographic dual geometry later.

\vspace{1cm}

\section{Wilson loops at Large `t Hooft Coupling} \label{asymptotic}

\vspace{5mm}
We have shown that evaluation of the Wilson loop $\langle W[C] \rangle$
is reduced to a related problem in a one-Hermitian matrix model.
Still, the matrix model is too complicated to solve exactly.
In the following, we focus our attention to either the ${\cal N}=2$ superconformal gauge theory of $A_1$ type with $G=$U$(N)$ coupled to $2N$ fundamental hypermultiplets and of $\hat{A}_1$ type with $G=$U$(N)\times$U$(N)$, both at large $N$ limit.
For these theories, we show that the large `t Hooft coupling behavior is determinable by a few quantities extracted from the one-loop determinant. This allows us to exactly evaluate the Wilson loop $\langle W[C] \rangle$ in the large $N$ and large 't Hooft coupling limit.

\vspace{5mm}

\subsection{General results in one matrix model} \label{generalMM}

\vspace{5mm}

Consider a matrix model for an $N\times N$ Hermitian matrix $X$.
In the large $N$ limit, expectation value of any operator in this model is determinable in terms of eigenvalue density function $\rho(x)$ of the matrix $X$.
By definition, $\rho(x)$ is normalized by
\begin{equation}
\int \rmd x\, \rho(x) = 1.
\end{equation}
Let $D$ denote the support of $\rho(x)$.
We assume that\footnote{
If $X$ is traceless, the assumption is always valid since $\int \rmd x\, \rho(x) \, x =0$ must hold.
In the large $N$ limit, the contribution from the trace part is negligible. }
\begin{equation}
\min\{ D \} =: b < 0 < a := \max\{ D \}.
\end{equation}

Expectation value of the operator $\frac1N\mbox{Tr}(e^{cX})$ $(c>0)$ is given in terms of $\rho(x)$ as
\begin{eqnarray}
W
&:=& \left\langle \frac1N\mbox{Tr}(e^{cX}) \right\rangle \nonumber \\
&=& \int \rmd x\,\rho(x)\, e^{cx}.
\end{eqnarray}
By the assumption on the support $D$, the value of $W$ is bounded:
\begin{equation}
e^{cb} \le W\le e^{ca}.
\end{equation}

\begin{figure}[ht!]
\vskip1cm
\centering
\includegraphics[scale=0.68]{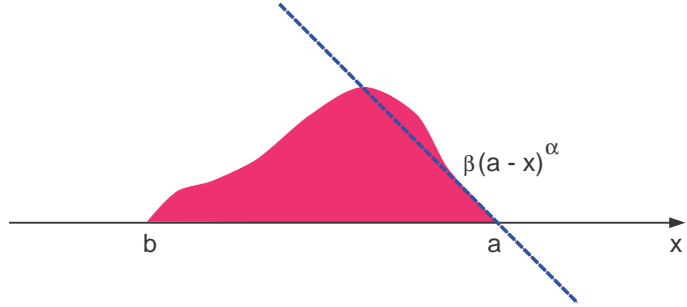}
\caption{\small \sl
Typical distribution of the eigenvalue density $\rho$.}
\label{}
\vskip1cm
\end{figure}

We are interested in the behavior of $W$ in the limit $a\to+\infty$.
Introducing the rescaled density function $\tilde{\rho}(x)=a\rho(ax)$, $W$ is written as
\begin{equation}
W = e^{ca}\int_0^{1-\frac ba}\rmd u\,\tilde{\rho}(1-u)e^{-cau} \qquad \mbox{where} \qquad
x=a(1-u).
\end{equation}
At the right edge of the support $D$, we expect that the density cuts off with a power-law tail:
\begin{equation}
\tilde{\rho}(1-u) = \beta u^\alpha+\chi(u) \qquad \mbox{where} \qquad |\chi(u)|\le Ku^{\alpha+\epsilon}, \hspace{5mm}
 u\in (0,\delta)
   \label{AE}
\end{equation}
for a positive $K,\epsilon,\delta$. See figure 2. Here, $\alpha > 0$ signifies the leading power of the fall-off at the right edge: $\chi$ refers to the sub-leading remainder. Then, for a large positive $a$, (\ref{AE}) leads to the following asymptotic behavior:
\begin{equation}
W \sim \beta\Gamma(\alpha+1)(ca)^{-\alpha-1}e^{ca},
   \label{estimate}
\end{equation}
Details of the derivation of (\ref{estimate}) are relegated to Appendix \ref{AEestimate}.

We have found that the large $a$ behavior of $W$ is determined by the functional
form of $\rho(x)$ in the vicinity of the right edge of its support.
In particular, we found that the leading exponential part is determined solely by the location of the right edge of the eigenvalue distribution.

\vspace{5mm}

For comparison, let us recall the exact form of the Wilson loop in ${\cal N}=4$ super Yang-Mills theory \cite{Erickson:2000af}, which is a special case of the $\hat{A}_0$ gauge theory.
In this case, the eigenvalue density function is given by
\begin{equation}
\rho(x) = \frac{4\pi}{\lambda}\sqrt{\frac{\lambda}{2\pi^2}-x^2} \ ,  \label{gaussiandensity}
\end{equation}
which is the solution of the saddle-point equation
\begin{equation}
\frac{4\pi^2}{\lambda}\phi = \int\hspace{-3.5mm}-\hspace{2mm}d\phi'\frac{\rho(\phi')}{\phi-\phi'}.
   \label{SYM}
\end{equation}
The Wilson loop is evaluated as follows:
\begin{eqnarray}
\langle W[C] \rangle
&=& \frac{4\pi}{\lambda}\int_{-\sqrt{\lambda}/\pi}^{+\sqrt{\lambda}/\pi} dx\,e^{2\pi x}\sqrt{\frac{\lambda}{2\pi^2}-x^2}
    \nonumber \\
&=& \frac2{\sqrt{2\lambda}}I_1(\sqrt{2\lambda})  \nonumber \\
&\sim& \sqrt{\frac2\pi}(2\lambda)^{-\frac34}e^{\sqrt{2\lambda}}.
\end{eqnarray}
We see that this asymptotic behavior is reproduced exactly by (\ref{estimate}) with $\alpha = {1 \over 2}$
of (\ref{gaussiandensity}) \footnote{
Here, the definition of the gauge coupling constant $g$ is different by the factor 2 from that in \cite{Erickson:2000af}
}.

\vspace{5mm}

\subsection{One-loop determinant and zeta function regularization} \label{subseczeta}

\vspace{5mm}

Let us return to the evaluation of $\langle W[C] \rangle$.
To determine the eigenvalue density function $\rho$ of the Hermitian matrix $\Phi$, it is necessary to know the explicit functional form of the one-loop determinant. However, this is a formidable task for a generic ${\cal N}=2$ gauge theory. Fortunately, as shown in the previous subsection, the leading behavior of $\langle W[C] \rangle$ is governed by a small number of data if $a=\mbox{max}\,(D)$ is large.

So, we shall assume that the limit $\lambda\to+\infty$ induces indefinite growth of $a$. This is a reasonable assumption since otherwise $\langle W[C] \rangle$ does not grow exponentially in the limit $\lambda\to+\infty$, implying that any ${\cal N}=2$ gauge theory with such a behavior of the Wilson loop cannot have an AdS dual in the usual sense.
In other words, we assume that the rescaled density function $\lambda^\gamma\rho(\lambda^\gamma x)$ has a
reasonable large $\lambda$ limit for a {\sl positive} $\gamma$.
Under this assumption, we now show that the large $\lambda$ behavior of the Wilson loop is determined by the behavior of the one-loop determinant in the region where the eigenvalues of $\Phi$ are large.
The asymptotic behavior in such a limit is most transparently derivable from the heat-kernel expansion for a certain differential operator in the zeta-function regularization of the one-loop determinant.

\vspace{5mm}
\noindent
$\bullet$ {\sl $A_1$ gauge theory}: \hfill\break
Consider first the $A_1$ gauge theory. There are contributions to the one-loop effective action both from the hypermultiplet and the vector multiplet. We first focus on the hypermultiplet contribution. If $\mathfrak{Q} \ V$ is expanded around the vanishing locus (\ref{locus}), quadratic terms of the hypermultiplet scalars become:
\begin{equation}
-q_\alpha(\Delta)^\alpha{}_\beta q^\beta+\frac1{r^2}\Phi^A\Phi^Bq_\alpha T_AT_Bq^\alpha,
   \label{quad_boson}
\end{equation}
where
\begin{eqnarray}
(\Delta)^\alpha{}_\beta
&=& (\nabla_m\delta^\alpha_\gamma+V_m{}^\alpha{}_\gamma)(\nabla^m\delta^\gamma_\beta+V^{m\gamma}{}_\beta)
    -\frac1{4r^2}(3+\cos^2\theta)\delta^\alpha_\beta, \\
V_m{}^\alpha{}_\beta &=& \bar{\xi}\Gamma^0{}_m\gamma^\alpha{}_\beta\tilde{\xi}.
\end{eqnarray}
If $\Phi$ is diagonalized as $\Phi=\mbox{diag}(\phi_1,\cdots,\phi_N)$, then the second term in (\ref{quad_boson}) can be written
as
\begin{equation}
\frac {2N}{r^2}\sum_{i=1}^N(\phi_i)^2q_{i\alpha}q^\alpha_i.
\end{equation}
Now the quadratic terms are decomposed into the sum of terms for components $q_i^\alpha$. So, the one-loop determinant of the hypermultiplet scalars is the product of determinants for each components.
Let $F_h^B(\Phi)$ denote a part of the matrix model action induced by the one-loop determinant for the hypermultiplet scalars $q^\alpha$. Its contribution to the effective action can be written as
\begin{equation}
F_h^B(\Phi) = 2N\sum_{i=1}^NF_h^B(\phi_i),
\end{equation}
where $F_h^B(m)$ is formally given as
\begin{equation}
F_h^B(m) := \log\mbox{Det}\Bigl( -\Delta+\frac{m^2}{r^2} \Bigr).
   \label{formaldef}
\end{equation}
Notice that the eigenvalues $\phi_i$ enter as masses of $q_i^\alpha$.
Therefore, what we need to analyze is the large $m$ behavior of $F_h^B(m)$.

We now evaluate the function $F_h^B(m)$ in the limit $m \rightarrow \infty$. In terms of Feynman
diagrammatics, this amounts to expanding the one-loop determinant in the background of scalar field $(m/r)^2$. Let $D(m)=\mbox{Det}(-\Delta+m^2/r^2)$.
The relation (\ref{formaldef}) is afflicted by ultraviolet infinities, so it should be regularized appropriately. The determinant is formally defined over the space spanned by the normalizable
eigenfunctions of $-\Delta$. Let $\lambda_k$ $(k=0,1,2,\cdots)$ be eigenvalues of $-\Delta$:
\begin{equation}
-\Delta\psi_k = \lambda_k\psi_k.
\end{equation}
Then, $D(m)$ can be formally written as
\begin{equation}
D(m) = \prod_{k=0}^\infty\Bigl( \lambda_k+\frac{m^2}{r^2} \Bigr).
\end{equation}
To make this expression well-defined, let us define a regularized function
\begin{equation}
\zeta(s,m) := r^{-2s}\sum_{k=0}^\infty\frac1{(\lambda_k+m^2/r^2)^s},  \label{regularized}
\end{equation}
where $s$ is a complex variable.
This summation may be well-defined for $s$ with sufficiently large Re$(s)$.
One can formally differentiate $\zeta(s,m)$ with respect to $s$ to obtain
\begin{equation}
\partial_s \zeta(s,m)\Big|_{s=0} = -\sum_{k=0}^\infty \log( r^2\lambda_k+m^2 ) = -\log [r^2D(m)].
\end{equation}
Since the left-hand side makes sense via a suitable analytic continuation of (\ref{regularized}), it can be regarded that the right-hand side is defined by the left-hand side. Therefore, we define the function
$F_h^B(m)$ via the zeta-function regularization:
\begin{equation}
F_h^B(m) := -\partial_s\zeta(s,m)\Big|_{s=0}.
\end{equation}

The large $m$ behavior of $F_h^B(m)$ is determined as follows. For a suitable range of $s$, $\zeta(s,m)$ can be written as
\begin{equation}
\zeta(s,m) = \frac{r^{-2s}}{\Gamma(s)}\int_0^\infty \rmd t\, t^{s-1}e^{-m^2t/r^2}K(t),
   \label{zeta}
\end{equation}
where
\begin{equation}
K(t) := \sum_{k=0}^\infty e^{-\lambda_kt} = \mbox{Tr}(e^{t\Delta})
\end{equation}
is the heat-kernel of $\Delta$. The convergence of this sum is assumed.
The asymptotic expansion of $K(t)$ is known as the heat-kernel expansion.
For a review on this subject, see e.g. \cite{Vassilevich:2003xt}.
Since $\Delta$ is a differential operator on $\mathbb{S}^4$, the heat-kernel expansion has the form
\begin{equation}
K(t) \sim \sum_{i=0}^\infty t^{i-2}a_{2i}(\Delta)
   \label{expansion}
\end{equation}
In the expansion, $a_{2i}(\Delta)$ are known as the heat-kernel coefficients for $\Delta$.

The expression (\ref{zeta}) of $\zeta(s,m)$ is only valid for a range of $s$, but $\zeta(s,m)$ can be analytically continued to the entire complex plane provided that the asymptotic expansion (\ref{expansion}) is known.
In particular, there exists a formula for the asymptotic expansion of $\zeta(s,m)$ in the large $m$ limit \cite{Voros:1986vw}
\begin{equation}
\zeta(s,m) \sim \sum_{i=0}^\infty a_{2i}(\Delta)r^{2i-4}\frac{\Gamma\left( s+i-2 \right)}{\Gamma(s)}m^{-2s-2i+4},
\end{equation}
valid in the entire complex $s$-plane.
Note that $a_{2i}(\Delta)r^{2i-4}$ are dimensionless combinations.
Differentiating with respect to $s$ and setting $s=0$, one obtains
\begin{eqnarray}
F_h^B(m)
&=& \Bigl( \frac12m^4\log m^2-\frac34m^4 \Bigr)a_0(\Delta)r^{-4}-\Bigl( m^2\log m^2-m^2 \Bigr)a_2(\Delta)r^{-2} \nonumber \\
& & +\log m^2\ a_4(\Delta)+O(m^{-2}\log m).
\end{eqnarray}

The evaluation of the one-loop determinant for the hypermultiplet fermions can be done similarly.
The quadratic terms of the fermions are given by
\begin{equation}
i\bar{\psi}\Gamma^m\nabla_m\psi-\frac ir\bar{\psi}\Gamma^0\Phi^AT_A\psi
 +\frac i2(\bar{\xi}\Gamma_{\mu\nu}\tilde{\xi})\bar{\psi}\Gamma^0\Gamma^{\mu\nu}\psi.
\end{equation}
We need to evaluate $-\log\mbox{Det}(iD\hspace*{-2.5mm}/\hspace{1mm})$ where
\begin{equation}
iD\hspace*{-2.5mm}/\hspace{1mm} := i\Gamma^m\nabla_m-\frac mri\Gamma^0+\frac \kappa2(\bar{\xi}\Gamma_{\mu\nu}\tilde{\xi})
 \Gamma^0\Gamma^{\mu\nu}
\end{equation}
with $\kappa=i$.
In the following, we will evaluate $-\frac12\log\mbox{Det}(iD\hspace*{-2.5mm}/\hspace{1mm})^2$ with a real $\kappa$, for which
$(iD\hspace*{-2.5mm}/\hspace{1mm})^2$ is non-negative and its heat-kernel is well-defined, and then
substitute $\kappa=i$ into the final expression. The validity of this procedure is justified by convergence of the result.

The explicit form of $(iD\hspace*{-2.5mm}/\hspace{1mm})^2$ is given by
\begin{eqnarray}
(iD\hspace*{-2.5mm}/\hspace{1mm})^2
&=& -(\nabla_m+{\cal V}_m)(\nabla^m+{\cal V}^m)-\frac12\Gamma^{mn}[\nabla_m,\nabla_n]-\frac{3\kappa^2}{4r^2}\sin^2\theta
    \nonumber \\
& & -\frac{\kappa^2}4(\bar{\xi}\Gamma_{\mu\nu}\tilde{\xi})(\bar{\xi}\Gamma_{\rho\sigma}\tilde{\xi})
    \Gamma^{\mu\nu}\Gamma^{\rho\sigma}
    +i\kappa\frac mr(\bar{\xi}\Gamma_{\mu\nu}\tilde{\xi})\Gamma^{\mu\nu}+\frac{m^2}{r^2} \nonumber \\
&:=& -\Delta_F+\frac{m^2}{r^2}.
\end{eqnarray}
where
\begin{equation}
{\cal V}_m = i\kappa(\bar{\xi}\Gamma_{m\mu}\tilde{\xi})\Gamma^0\Gamma^\mu.
\end{equation}

The fermion case is slightly different from the scalar case since there is a term linear in $m$ in $-\Delta_F$.
However, the asymptotic expansion of the zeta-function-regularized one-loop determinant can be made in the fermion case as well.
The part $F_h^F(\Phi)$ of the matrix model action due to $\psi$ has a similar form with $F_h^B(\Phi)$, with different coefficients.

The total one-loop contribution of hypermultiplet to the effective action is $F_h=F_h^B+F_h^F$.
Because of underlying supersymmetry, the terms of order $m^4$ and $m^4\log m^2$ cancel between $F_h^B$ and $F_h^F$. The resulting expression for $F_h$ is
\begin{eqnarray}
F_h &=& 2N\sum_{i=1}^NF(\phi_i), \\
F(m) &=& c_1m^2\log m^2+c_2m^2+c_3\log m^2+O(m^{-2}\log m).
   \label{F(m)}
\end{eqnarray}
The fact that $c_1$ is positive will turn out to be important later, while the exact values of the coefficients are irrelevant for the large `t Hooft coupling behavior of the Wilson loop.
We presented details of computation of $c_1$ in Appendix \ref{coeff}.
Notice that, at least up to this order, $F(m)$ is an even function of $m$.

\vspace{5mm}
Obviously, $F_h$ depends on field contents.
The expression for $F_h$ when $R$ is the adjoint representation can be found easily by noticing that, for example, the 'mass' term of $q^\alpha$ can be put to
\begin{equation}
\frac1{r^2}\sum_{i\ne j}(\phi_i-\phi_j)^2q_{ij\alpha}q^{\alpha}_{ji}.
\end{equation}
In this case, $F_h$ is written as
\begin{eqnarray}
F_h \Big\vert_{\rm adj.} = \sum_{i\ne j}F(\phi_i-\phi_j).
\label{f-adj}
\end{eqnarray}
Note that $F(m)$ here is the same function as (\ref{F(m)}).

Direct evaluation of the contribution from the vector multiplet, which we denote as $F_v$, appears more complicated since there are mixing terms between $A_m$ and $A_a$.
Fortunately, it was shown in \cite{Pestun:2007rz} that $F_v$ and $F_h$ cancel each other in ${\cal N}=4$ super Yang-Mills theory. This implies from (\ref{f-adj}) that
\begin{equation}
F_v = -\sum_{i\ne j}F(\phi_i-\phi_j).
\end{equation}

\noindent
$\bullet$ {\sl $\hat{A}_1$ gauge theory}:\hfill\break
We next consider the $\hat{A}_1$ quiver gauge theory. In this case,  $q^\alpha$ and $\psi$ consist of bi-fundamental fields. The $\Phi$ is a block-diagonal matrix:
\begin{equation}
\Phi = \left(
\begin{array}{cc}
\Phi^{(1)} & \\ & \Phi^{(2)}
\end{array}
\right),
\end{equation}
in which $\Phi^{(1)}=\mbox{diag}(\phi^{(1)}_1,\cdots,\phi^{(1)}_N)$ and $\Phi^{(2)}=\mbox{diag}(\phi^{(2)}_1,\cdots,\phi^{(2)}_N)$, respectively.
By repeating the similar computations, one can easily show that $F_h$ has the form
\begin{equation}
F_h = 2\sum_{i, j=1}^NF(\phi^{(1)}_i-\phi^{(2)}_j),
\end{equation}
and $F_v$ has the form
\begin{equation}
F_v = -\sum_{i\ne j}F(\phi^{(1)}_i-\phi^{(1)}_j)-\sum_{i\ne j}F(\phi^{(2)}_i-\phi^{(2)}_j).
\end{equation}
The total one-loop contribution is the sum $F = F_h + F_v$.

As a consistency check of the above result,
consider taking the two nodes identical. This reduces the number of nodes from two to one, and hence must map the $\hat{A}_1$ gauge theory to $\hat{A}_0$ one. The reduction puts $\Phi^{(1)}$ and $\Phi^{(2)}$ equal. Then, up to an irrelevant constant, $F_v$ is precisely minus of $F_h$. We thus see that $F$ vanishes identically, reproducing the known result of the $\hat{A}_0$ gauge theory.

\vspace{5mm}

\subsection{Saddle-point equations} \label{saddlepoint}

\vspace{5mm}

We can now extract the saddle-point equations for the matrix model and determine the large `t Hooft coupling behavior of the Wilson loop from them. \hfill\break
\vskip0.3cm
\noindent
$\bullet$ {\sl $A_1$ gauge theory}: \hfill\break
In this theory, the saddle-point equation reads
\begin{equation}
\frac{8\pi^2}{\lambda}\phi_k+2F'(\phi_k)-\frac2N\sum_{i\ne k}F'(\phi_k-\phi_i) = \frac2N\sum_{i\ne k}\frac1{\phi_k-\phi_i}.
\label{a1saddle}
\end{equation}
As explained before, we assume that $\lambda^\gamma\rho(\lambda^\gamma \phi)$ for a {\sl positive} $\gamma$ has a sensible large $\lambda$ asymptote. By rescaling $\phi_k\to\lambda^\gamma\phi_k$, one obtains
\begin{equation}
8\pi^2\phi_k+2\lambda^{1-\gamma}F'(\lambda^\gamma\phi_k)-\frac2N\sum_{i\ne k}\lambda^{1-\gamma}F'(\lambda^\gamma(\phi_k-\phi_i))
 = \frac2N\lambda^{1-2\gamma}\sum_{i\ne k}\frac1{\phi_k-\phi_i}.
\end{equation}
Recall that $F(x)\sim c_1x^2\log x^2$ for large $x$.
This shows that the leading-order equation for large $\lambda$ is given by
\begin{equation}
4c_1\phi_k\log \phi_k+2(c_1+c_2)\phi_k-\frac2N\sum_{i\ne k}\Bigl[ 2c_1(\phi_k-\phi_i)\log (\phi_k-\phi_i)+(c_1+c_2)
 (\phi_k-\phi_i) \Bigr] = 0.
\end{equation}
Differentiating twice with respect to $\phi_k$, one obtains
\begin{equation}
\frac1{\phi_k} = \frac1N\sum_{i\ne k}\frac1{\phi_k-\phi_i}.
\end{equation}
Notice that $c_1$ and $c_2$ dropped out. Now, this equation has no sensible solution.
Therefore, we conclude that the scaling assumption we started with is invalid, implying that the Wilson loop in this theory cannot grow exponentially in the large `t Hooft coupling limit.
%Replacing $\lambda^\gamma$ with $a$, the right edge of the eigenvalue distribution,
%the same argument implies that the Wilson loop is finite in the limit.

There is another way to check the finiteness of the Wilson loop.
Let us rewrite the saddle-point equation as follows:
\begin{equation}
\frac{8\pi^2}{\lambda}\phi_k+2F'(\phi_k) = \frac2N\sum_{i\ne k}F'(\phi_k-\phi_i) + \frac2N\sum_{i\ne k}\frac1{\phi_k-\phi_i}.
\label{a1saddle}
\end{equation}
The left-hand side represents the external force acting on the eigenvalues, while the right-hand side represents the interactions among the eigenvalues.
For a large $\phi_k$, the external force is dominated by $2F'(\phi_k)$, which is nonzero.
This implies that the large $\lambda$ limit must be smooth, and the Wilson loop expectation value approaches a finite value.
Recall that in the case of ${\cal N}=4$ super Yang-Mill theory, the large $\lambda$ limit renders the external force to vanish, resulting in an indefinite spread of the eigenvalues. This is reflected in the exponential growth of the Wilson loop expectation value.

Implications of this surprising conclusion are far reaching: the ${\cal N}=2$ supersymmetric gauge theory coupled to $2N$ fundamental hypermultiplets, although superconformal, must have a holographic dual whose geometry does not belong to the more familiar cases such as ${\cal N}=4$ super Yang-Mills theory. Central to this phenomenon is that there are two `t Hooft coupling parameters whose ratio can be tuned hierarchically large or small. In particular, we can tune one of them to be smaller than ${\cal O}(1)$, which also renders two widely separated length scales (in units of string scale) in the putative gravity dual background.
In the next section, we shall discuss how nonstandard the dual geometry ought to be by using the non-exponential behavior of the Wilson loop as a probe.

\vspace{3mm}
\noindent
$\bullet$ {\sl $\hat{A}_1$ gauge theory}: \hfill\break
In this theory,
there are two saddle-point equations corresponding to two matrices $\Phi^{(1)}$ and $\Phi^{(2)}$:
\begin{eqnarray}
\frac{8\pi^2}{\lambda_1}\phi^{(1)}_k+\frac2N\sum_{i=1}^NF'(\phi^{(1)}_k-\phi^{(2)}_i)
 -\frac2N\sum_{i\ne k}F'(\phi^{(1)}_k-\phi^{(1)}_i) = \frac2N\sum_{i\ne k}\frac1{\phi^{(1)}_k-\phi^{(1)}_i}, \\
\frac{8\pi^2}{\lambda_2}\phi^{(2)}_k+\frac2N\sum_{i=1}^NF'(\phi^{(2)}_k-\phi^{(1)}_i)
 -\frac2N\sum_{i\ne k}F'(\phi^{(2)}_k-\phi^{(2)}_i) = \frac2N\sum_{i\ne k}\frac1{\phi^{(2)}_k-\phi^{(2)}_i},
\label{discretesaddlepointeqn}
\end{eqnarray}
where $\lambda_1=g_1^2N$ and $\lambda_2=g_2^2N$ are the `t Hooft coupling constants of each gauge groups.

Denote $\rho^{(1)}(\phi)$, $\rho^{(2)}(\phi)$ the eigenvalue distribution functions for the $\Phi^{(1)}$, $\Phi^{(2)}$ matrices, respectively. It is convenient to define
\begin{eqnarray}
\rho(\phi) &:=& \frac12(\rho^{(1)}(\phi)+\rho^{(2)}(\phi)), \\
\delta\rho(\phi) &:=& \frac12(\rho^{(1)}(\phi)-\rho^{(2)}(\phi)).
\end{eqnarray}
In terms of them, the above saddle-point equations are simplified as follows:
\begin{eqnarray}
\frac{4\pi^2}{\lambda}\phi &=& \int\hspace{-4mm}-\hspace{2.5mm}\rmd\phi'\frac{\rho(\phi')}{\phi-\phi'},
   \label{A_2} \label{untwisted} \\
2\pi^2\Bigl[ \frac1{\lambda_1}-\frac1{\lambda_2} \Bigr]\phi
 -2\int\hspace{-4mm}-\hspace{2.5mm}\rmd\phi'\delta\rho(\phi')F'(\phi-\phi')
 &=& \int\hspace{-4mm}-\hspace{2.5mm}\rmd\phi'\frac{\delta\rho(\phi')}{\phi-\phi'}, \label{twisted}
\end{eqnarray}
where
\begin{equation}
\frac1{{\lambda}} := {1 \over |\Gamma|} \left( \frac1{\lambda_1}+\frac1{\lambda_2} \right) \qquad \mbox{and} \qquad |\Gamma| = 2.
   \label{lambdaforA_2}
\end{equation}
For obvious reasons, we refer these two as untwisted and twisted saddle-point equations.
By the scaling argument, one can show that $\delta\rho(\phi)$ is negligible compared to $\rho(\phi)$
in the large $\lambda$ limit. In particular, when $\lambda_1 = \lambda_2$, it follows that $\delta \rho = 0$ is a solution, consistent with $\mathbb{Z}_2$ parity exchanging the two nodes. Therefore, the large $\lambda$ behavior of the Wilson loop is determined by (\ref{A_2}), which is exactly the same as (\ref{SYM}). Indeed, $\lambda$ defined by (\ref{lambdaforA_2}) is exactly what is related to $g_sN$ \cite{Klebanov:1999rd}.

The two Wilson loops are then obtainable from the one-matrix model with eigenvalue density $\rho \pm \delta \rho$:
\bea
&& W_{1} = \int_D \rmd x \, e^{a x} \rho^{(1)}(x)\ = \ \int_D \rmd x \, e^{a x} [\rho(x) + \delta \rho(x)] \nonumber \\
&& W_{2} = \int_D \rmd x \, e^{a x} \rho^{(2)}(x) \ = \ \int_D \rmd x \, e^{a x} [\rho(x) - \delta \rho(x)]. \label{twowilsonloops}
\eea
We see that the untwisted and the twisted Wilson loops are given by
\bea
&& W^{(0)} := {1 \over 2} (W_1 + W_2) = \int_D \rmd x \, e^{a x} \, \rho (x) \nonumber \\
&& W^{(1)}:= {1 \over 2} (W_1 - W_2) = \int_D \rmd x \, e^{a x} \, \delta \rho (x).
\eea
Inferring from the saddle-point equations (\ref{untwisted}, \ref{twisted}), we see that these Wilson loops are directly related to the
average and difference of the two gauge coupling constants. It also shows that the twisted Wilson loop will have nonzero expectation value once the two gauge couplings are set different. In the next section, we shall see that they descend from moduli parameters of six-dimensional twisted sectors at the orbifold singularity in the holographic dual description.

We have found the following result for the Wilson loop in $\hat{A}_1$ quiver gauge theory. The two Wilson loops, corresponding to the two quiver gauge groups, have exponentially growing behavior at large `t Hooft coupling limit. Its functional form is exactly the same as the one exhibited by the Wilson loop in ${\cal N}=4$ super Yang-Mills theory.

\vskip0.5cm
\subsection{Interpolation among the quivers}
With the saddle-point equations at hand, we now discuss various interpolations among $\hat{A}_0, A_1, \hat{A}_1$ theories and learn about the gauge dynamics. Our starting point is the $\hat{A}_1$ theory, whose quiver diagram has two nodes. See figure 1.
\hfill\break
\vskip0.3cm
\nonumber
$\bullet$ Consider the symmetric quiver for which the two `t Hooft coupling constants take the ratio $\lambda_1 / \lambda_2 = 1$. Then the twisted saddle-point equation (\ref{twisted}) asserts that $\delta \rho = 0$ is the solution. It follows that $\langle W_1 \rangle - \langle W_2 \rangle = 0$, viz. the Wilson loop in the twisted sector vanishes identically. Intuitively, the two gauge interactions are of equal strength, so the two Wilson loops are indistinguishable.
Moreover, from the untwisted saddle-point equation (\ref{untwisted}),  we see that the Wilson loop in the untwisted sector behaves exactly the same as the one in $\hat{A}_0$ theory and, in particular, ${\cal N}=4$ super Yang-Mills theory:
\bea
W^{(0)} = {1 \over 2} \Big(\langle W_1 \rangle + \langle W_2 \rangle \Big) = {1 \over \sqrt{2 \lambda}} I_1 (\sqrt{2 \lambda}).  \label{untwistedWilson}
\eea
It follows that the Wilson loop grows exponentially at large `t Hooft coupling limit, much the same way as the $\hat{A}_0$ theory does.
\hfill\break
\vskip0.3cm
\nonumber
$\bullet$ Consider the asymmetric quiver where the ratio $\lambda_1 / \lambda_2 \ne 1$ but finite. The twisted saddle-point equation (\ref{twisted}) can be recast as
\bea
{1 \over \lambda} \left( B - {1 \over 2} \right) \int\hspace{-4mm}-\hspace{2.5mm} \rmd \phi' {\rho(\phi') \over \phi - \phi'} = \int\hspace{-4mm}-\hspace{2.5mm} \rmd \phi' \, \delta \rho (\phi') \left[{1 \over 2} {1 \over \phi - \phi'}+ F'(\phi - \phi')\right]. \label{twisted2}
\eea
Here, we parametrized the difference of two inverse `t Hooft couplings as
\bea
\left(B - {1 \over 2} \right) := {1 \over 2} \left({1 \over \lambda_1} - {1 \over \lambda_2} \right) \Big/ \left({1 \over \lambda_1} + {1 \over \lambda_2}\right).
\eea
Obviously, taking into account the $\mathbb{Z}_2$ exchange symmetry between the two quiver nodes, $B$ ranges over the interval $[0, +1]$. The symmetric quiver considered above corresponds to $B = {1 \over 2}$. Solving first $\rho$ from (\ref{untwisted}) and substituting the solution to (\ref{twisted2}), one solves $\delta \rho$ as a function of $B$. We see from (\ref{twisted2}) that $\delta \rho$ ought to be a {\sl linear} function of $B$ throughout the interval $[0, +1]$. Equivalently, extending the range of $B$ to $(-\infty, +\infty)$, we see that $\delta \rho$ is a sawtooth function, piecewise linear over each unit interval of $B$. In particular, it is discontinuous across $B=0$ (and across all other nonzero integer values). This is depicted in figure 3. Therefore, we conclude that the Wilson loops $W_1, W_2$ at strong `t Hooft coupling limit are nonanalytic not only in $\lambda$ but also in $B$. In fact, as we shall recall in the next section, $B=0$ is a special point where the spacetime gauge symmetry is enhanced and the worldsheet conformal field theory becomes singular. Nevertheless, the Wilson loop in the untwisted sector behaves exactly the same as the symmetric quiver, viz. (\ref{untwistedWilson}). We conclude that the untwisted Wilson loop is independent of strength of the gauge interactions.

\begin{figure}[ht!]
\vskip1cm
\centering
\includegraphics[scale=0.68]{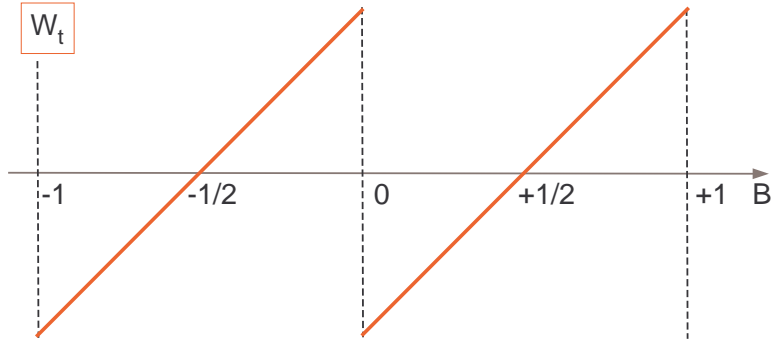}
\caption{\small \sl
Dependence of twisted sector Wilson loops on the parameter $B$. It shows discontinuity at $B=0$, resulting in non-analytic behavior of the Wilson loops to both gauge couplings. }
\label{}
\vskip1cm
\end{figure}

\hfill\break
\vskip0.3cm
$\bullet$ Consider an extreme limit of the asymmetric quiver where the ratio $\lambda_1/\lambda_2 \rightarrow 0$, equivalently, $\lambda_2/\lambda_1 \rightarrow \infty$, viz. the two `t Hooft couplings are hierarchically separated. In this case, one gauge group is infinitely stronger than the other gauge group and the $\hat{A}_1$ quiver gauge theory ought to become the $A_1$ gauge theory
%\footnote{Considerations similar to this was studied systematically in \cite{gaiotto}.}
. This can be seen as follows. In the $\hat{A}_1$ saddle-point equations (\ref{discretesaddlepointeqn}), we see that $\phi^{(1)} \rightarrow 0$ solves the first equation.
Plugging this into the second equation, we see it is reduced to the $A_1$ saddle-point equation (\ref{a1saddle}). This reduction poses a very interesting physics since from the above considerations the Wilson expectation value interpolates from the exponential growth of the $\hat{A}_1$ quiver gauge theory to the non-exponential behavior of the $A_1$ gauge theory. In the next section, we shall argue that this is a clear demonstration (as probed by the Wilson loops) that holographic dual of the $A_1$ gauge theory ought to have internal geometry of {\sl string scale} size.

We can also understand the interpolation directly in terms of the Wilson loop. Consider, for example, $\lambda_2 / \lambda_1 \rightarrow \infty$. From the $\hat{A}_1$ Wilson
loops, using the fact that $\rho^{(1)}(x), \rho^{(2)}(x)$ are strictly positive-definite, we have
\bea
\langle W_{2} \rangle &=& \int \rmd \lambda \ \rho^{(2)}(\lambda) \ e^\lambda \nonumber \\
& \le & 2 \int \rmd \lambda \ {1 \over 2} [\rho^{(1)}(\lambda) + \rho^{(2)}(\lambda) ] e^\lambda \nonumber \\
&= & \frac4{\sqrt{2\lambda}}I_1(\sqrt{2\lambda}) \ .
\eea
Since $\lambda \sim \lambda_1 \to 0$, the Wilson loop is bounded from above by a constant.
Note that the limit $\lambda_1\to0$ can be safely taken: the saddle-point equation (\ref{untwisted}) is in fact exact in $\lambda$.
\vskip0.3cm
\noindent
$\bullet$ Consider the limit $\lambda_1, \lambda_2 \rightarrow 0$. In this limit,
\bea
\lambda = 2{ \lambda_1 \lambda_2 \over \lambda_1 + \lambda_2} \ \rightarrow \ 0 \ , \qquad
\kappa := {\lambda_2 \over \lambda_1} = {\rm fixed} \label{changeofvariables}
\eea
and the exact result (\ref{untwistedWilson}) is expandable in power series of $\lambda$ and $\kappa$:
\bea
W^{(0)} \Big\vert_{\rm exact} = {1 \over 2} \Big(\langle W_1 \rangle + \langle W_2 \rangle \Big)
%={1 \over 2} \sum_{\ell =0}^\infty {(2 \lambda)^\ell \over \ell! (\ell+1)!}
= {1 \over 2} \sum_{\ell=0}^\infty \sum_{m=0}^\infty {(-)^m (\ell + m - 1)! \over (\ell - 1)! \ell ! (\ell + 1)!}
 \lambda_1^\ell \kappa^{\ell + m} \ .  \label{weakcoupling}
\eea
Here, the exact result (\ref{untwistedWilson}) is symmetric under $\lambda_1 \leftrightarrow \lambda_2$, so we assumed in (\ref{weakcoupling}) that $\kappa < 1$. On the other hand, from standpoint of the quiver gauge theory, the Wilson loop in the fixed-order perturbation theory is given by power series in $\lambda_1$ or $\lambda_2$:
\bea
W^{(0)} \Big\vert_{\rm pert} = \sum_{\ell=0}^\infty \sum_{m=0}^\infty
W_{\ell, m} \lambda_1^\ell \lambda_2^m = \sum_{\ell = 1}^\infty \sum_{m=1}^\infty W_{\ell, m} \lambda_1^{\ell + m} \kappa^m. \label{weakcouplingexpansion}
\eea

We see that the exact result (\ref{weakcoupling}) and the perturbative result (\ref{weakcouplingexpansion}) do not agree each other. Recall that both results are obtained at planar limit $N \rightarrow$ and ought to be absolutely convergent in $(\lambda, B)$ and in $(\lambda_1, \lambda_2)$, respectively. The reason may be
that the two sets of coupling constants are not analytic in $\mathbb{C}^2$ complex plane. In fact, from (\ref{changeofvariables}), we see that $\lambda(\lambda_1, \lambda_2)$ has a codimension-1 singularity at $\lambda_1 + \lambda_2 = 0$. An exceptional situation is when $\lambda_1 = \lambda_2$. In this case, the
singularity disappears and, with the same power series expansion, we expect the exact result (\ref{weakcoupling}) and the perturbative result (\ref{weakcouplingexpansion}) are the same.

We should note that the change of variables is well-defined at strong coupling regime. In this regime, power series expansions in $1/\lambda_1$ and $1 / \lambda_2$ is related unambiguously to power series expansions in $1/\lambda$ and $B$. In fact, the change of variables
\bea
\Big( \ {1 \over \lambda_1}, {1 \over \lambda_2} \ \Big) \quad \longrightarrow \quad
\Big(\ {1 \over \lambda}, B \ \Big)
\eea
is analytic and does not introduce any singularity around $\lambda_1, \lambda_2 = \infty$. In fact, as we
will recapitulate, these are the variables naturally introduced in the gravity dual description.

We remark that the analytic structure of the Wilson loops in quiver gauge theories is similar to the Ising model in a magnetic field on a planar random lattice \cite{kazakov}. The latter is defined by a matrix model involving two interacting Hermitian matrices and involves two coupling parameters: average `t Hooft coupling and magnetic field. Here again, by turning on the magnetic field, one can scale two independent `t Hooft coupling parameters differently. In light of our results, it would be extremely interesting to study this
system in the limit the magnetic field is sent to infinity.

\vskip0.3cm
%%%%%%%%%%%%%%%%%%%%%%%%%%%%%%%%%%%%%%%%%%%%%%%%%%%%%%%%%%%%%%%%%%%%%%%%%%%%%%%%%%%%%%%%%%%%%%%%%%%%%%%%%
\section{Intuitive Understanding of Non-Analyticity}
In the last section, the distinguishing feature of the $A_1$ theory from the $\hat{A}_0, \hat{A}_1$ theories was that growth of the Wilson loop expectation value was less than exponential. Yet, these theories are connected one another by continuously deforming gauge coupling parameters. How can then such a non-analytic behavior come about? \footnote{This question was raised to us by Juan Maldacena.} In this section, we offer an intuitive understanding of this in terms of competition between screening and over-screening of color charges and also draw analogy to the Kondo effect of magnetic impurity in a metal.
\vskip0.5cm

\noindent $\bullet$ {\sl screening versus anti-screening}: \hfill\break
Consider first the weak coupling regime. The representation contents of these ${\cal N}=2$ quiver gauge theories are such that the $\hat{A}_0$ theory contains field contents in adjoint representations only, while the $\hat{A}_1$ and the $A_1$ theories contain additional field contents in bi-fundamental or fundamental representations, respectively. The $A_1$ theory contains additional massless multiplets in fundamental representation, so we see immediately that the theory is capable of screening an external color charge sourced by the Wilson loop for any representations. Since the theory is conformal, the screening length ought to be infinite (zero is also compatible with conformal symmetry, but it just means there is no screening) and impeding creation of an excitation energy above the ground state. Even more so, `tension' of the color flux tube would go to zero. In other words, once a static color charge is introduced to the theory, massless hypermultiplets in fundamental representation will immediately screen out the charge to arbitrary long distances. Though this intuitive picture is based on weak coupling dynamics, due to conformal symmetry, it fits well with the non-exponential growth of the Wilson loop in the $A_1$ theory, which we derived in the previous section in the planar limit.

We stress that the screening has nothing to do with supersymmetry but is a consequence of elementary consideration of gauge dynamics with massless matter in complex representations. This is clearly illustrated by the well known two-dimensional Schwinger model. Generalization of this Schwinger mechanism to nonabelian gauge theories showed that massless fermions in {\sl arbitrary} complex representation screens the heavy probe charge in the fundamental representation \cite{Gross}. The screening and consequent string breaking by the dynamical massless matter was observed convincingly in both two-dimensional QED \cite{screen2} and three-dimensional QCD \cite{screen3}. In four-dimensional lattice QCD, the static quark potential $V(R)a$ was computed ($a$ denotes the lattice spacing) for fermions in both quenched and dynamical simulations \cite{screen4}. For quenched simulation, the potential scaled linearly with $R/a$, indicating confinement behavior. For dynamical simulation, the potential exhibited flattening over a wide range of the separation distance $R/a$.

\begin{figure}[ht!]
\vskip1cm
\centering
\includegraphics[scale=0.68]{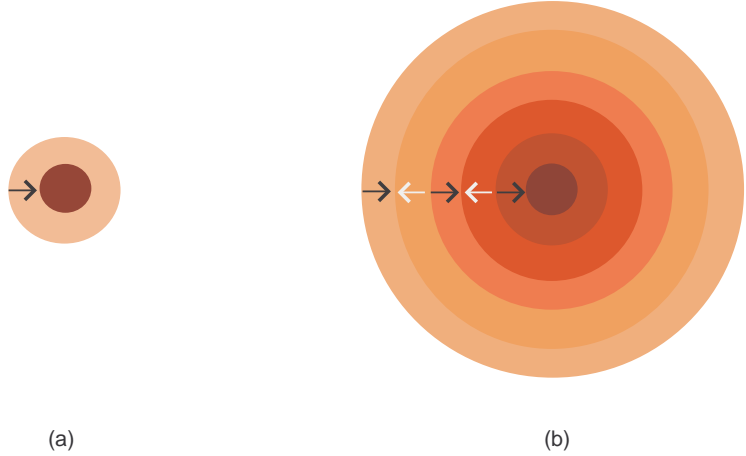}
\caption{\small \sl Response of gauge theories to external color charge source. (a) For $A_1$ theory, an external color charge in fundamental representation of the gauge group is screened by the $N_f = 2 N_c$ flavors of massless matter fields, which are in fundamental representation (blue arrow). (b) For $\hat{A}_1$ theory, an external color charge in fundamental representation of the first gauge group is screened by the massless matter fields. As the matter fields are in bi-fundamental representations (black and white arrows), color charge in the second gauge group is regenerated and anti-screened. The process repeats between the two gauge groups and leads the theory to exhibit Coulomb behavior.}
\label{}
\vskip1cm
\end{figure}

The case of $\hat{A}_1$ theory is more interesting. Having two gauge groups associated with each nodes, consider introducing a static color charge of the representation $R$ for, say, the first gauge group in SU$(N) \times $SU$(N)$. The hypermultiplets transforming in $({\bf N}, \overline{\bf N})$ and $(\overline{\bf N}, {\bf N})$ are in defining representations with respect to the first gauge group, so they will rearrange their ground-state configuration to screen out the color charge. But then, as these hypermultiplets are in defining representation with respect to the second gauge group as well, a complete screening with respect to the first gauge group will reassemble the resulting configuration to be in the representation $\overline{R}$ of the second gauge group in SU$(N)\times$SU($N$). This configuration is essentially the same as the starting configuration except that the two gauge groups are interchanged (along with charge conjugation). The hypermultiplets may opt to rearrange their ground-state configuration to screen out the color charge of the second gauge group, but then the process will repeat itself and returns back to the original static color charge of the first gauge group --- in $\hat{A}_1$ theory, perfect screening of the first gauge group is accompanied by perfect anti-screening of the second gauge group and vice versa. This is depicted in figure 4. Consequently, a complete screening never takes place for {\sl both} gauge groups simultaneously. Instead, the external color charge excites the ground-state to a conformally invariant configuration with the Coulomb energy. Again, we formulated this intuitive picture from weak coupling regime, but the picture fits well with the exponential growth of the Wilson loop expectation value of $\hat{A}_1$ theory we derived in the previous section at planar limit.
\vskip0.5cm

\noindent $\bullet$ {\sl Analogy to Kondo effect}: \hfill\break
It is interesting to observe that the screening vs. anti-screening process described above is reminiscent of the multi-channel Kondo effect in a metal \cite{affleck}. There, a static magnetic impurity carrying a spin $S$ interacts with conduction electrons and profoundly affects electrical transport property at long distances. Suppose in a metal there are $N_{\rm f}$ flavors of conduction band electrons. Thus, there are $N_{\rm f}$ channels and they are mutually non-interacting. The antiferromagnetic spin-spin interaction between the impurity and the conduction electrons leads at weak coupling to screening of the impurity spin $S$ to $S_{\rm ren} = (S - N_{\rm f}/2)$. We see that the system with $N_{\rm f} < 2 S$ is under-screened, leading to an asymptotic screening of the impurity spin and that the system with $N_{\rm f} > 2S$ is over-screened, leading to an asymptotic anti-screening of the impurity spin. The marginally screened case, $N_{\rm f} = 2 S$, is at the border between the screening and the anti-screening: the spin $S$ of the magnetic impurity is intact under renormalization by the conduction electrons (modulo overall flip of the spin orientation, which is a symmetry of the system). We thus observe that the Coulomb behavior of the external color source in $\hat{A}_1$ theory is tantalizingly parallel to the marginally screened case of the multi-channel Kondo effect.
\vskip0.5cm

\noindent $\bullet$ {\sl Interpretation via brane configurations}: \hfill\break
We can also understand the screening-Coulomb transition from the brane configurations describing $\hat{A}_1$ and $A_1$ theories \footnote{For a comprehensive review of brane configurations, see \cite{giveonkutasov}.}. Consider Type IIA string theory on $\mathbb{R}^{8,1} \times \mathbb{S}^1$, where the circle direction is along $x^9$ and have circumference $L$. We set up the brane configuration by introducing two NS5-branes stretched along $(012345)$ directions and $N$ stack of D4-branes stretched along $(01239)$ directions on intervals between the two NS5-branes. Generically, the two NS5-branes are located at separate position on $\mathbb{S}^1$ and this corresponds to the $\hat{A}_1$ theory. The gauge couplings $1/g_1^2$ and $1/g_2^2$ of the two quiver gauge groups are proportional to the length of the two $x^9$-intervals of the D4-branes. When the two NS5-branes are located at diagonally opposite points, say, at $x^9 = 0, L/2$, the two gauge couplings of the $\hat{A}_1$ theory are equal. This is depicted in figure 5(a). By approaching one
NS5-brane to another, say, at $x^9=0$, we can obtain the configuration in figure 5(b). This corresponds to $A_1$ theory since the gauge coupling of the D4-branes encircling the $\mathbb{S}^1$ becomes arbitrarily weak compared to that of the D4-branes stretched infinitesimally between the two overlapping NS5-branes.

\begin{figure}[ht!]
\vskip1cm
\centering
\includegraphics[scale=0.68]{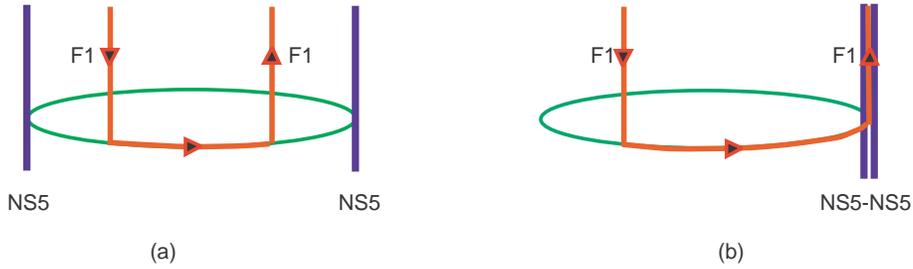}
\caption{\small \sl
Semiclassical Wilson loop in brane configuration of ${\cal N}=2$ superconformal gauge theories under study: (a) $\hat{A}_1$ theory with $G=$SU($N) \times$ SU($N$) and $2N$ bifundamental hypermultiplets. $N$ D4-branes stretch between two widely separated NS5-branes on a circle. The F1 (fundamental string) ending on or emanating from D4-brane represent static charges. On D4-branes, having finite gauge coupling, conservation of the F1 flux is manifestly. (b) $A_1$ theory with $G$=SU($N$) and $2N$ fundamental hypermultiplets. The $A_1$ theory is obtained from $\hat{A}_1$ in (a) by approaching the two NS5-branes. The flux is leaked into the coincident NS5-branes and run along their worldvolumes. On D4-branes, having vanishing gauge coupling, conservation of the F1 flux is not manifest. }
\label{}
\vskip1cm
\end{figure}

We now introduce external color charge to the D4-branes and examine fate of the color fluxes. The external color sources are provided by a macroscopic IIA fundamental string ending on the stacked D4-branes. Consider first the configuration of the $\hat{A}_1$ theory. The color charge is an endpoint of the fundamental string on one stack of the D4-branes, viz. one of the two quiver gauge groups. Along the D4-branes, the endpoint sources color Coulomb field. The color field will sink at another external color charge located at a finite distance from the first external charge. See figure 2(a). We see that the color flux is conserved on the first stack of D4-branes. We also see that, at weak coupling regime, effects of the NS5-branes are negligible.

Consider next the configuration of the $A_1$ theory. Based on the considerations of the previous section,
we consider an external color charge to the stack of D4-branes encircling the $\mathbb{S}^1$. In this configuration, the two NS5-branes are coincident and this opens up a new possible color flux configuration. To understand this, we recall the situation of stack of D1-D5 branes, which is related to the macroscopic IIA string and stack of NS5-branes. In the D1-D5 system, it is well known that there are threshold bound states of D1-branes on D5-branes {\sl provided} two or more D5-branes are stacked. For a single D5-brane, the D1-brane bound-state does not exist. This suggests in the brane configuration of the $A_1$ theory that
the color flux may now be pulled to and smear out along the two coincident NS5-branes. From the viewpoint
of stack of the D4-branes encircling $\mathbb{S}^1$, the color flux appears not conserved.

%%%%%%%%%%%%%%%%%%%%%%%%%%%%%%%%%%%%%%%%%%%%%%%%%%%%%%%%%%%%%%%%%%%%%%%%%%%%%%%%%%%%%%%%%%%%%%%%%%%%%%%%%
\section{Holographic Dual} \label{holography}

The exact results of the ${\cal N}=2$ Wilson loops at strong `t Hooft coupling limit we obtained in the previous section revealed many intriguing aspects. In particular, compared to the more familiar, exponential growth behavior of the ${\cal N}=4$ Wilson loops, we found the following distinguishing features and consequences:

\begin{list}{$\bullet$}{}

\item In $A_1$ gauge theory, the Wilson loop $\langle W \rangle$ does {\sl not} exhibit the exponential growth. Replacing $2N$ fundamental representation hypermultiplets by single adjoint representation hypermultiplet restores the exponential growth, since the latter is nothing but the ${\cal N}=4$ counterpart. This suggests that $\langle W \rangle$ in $\hat{A}_1$ gauge theory has (possibly infinitely) many saddle points and potential leading exponential growth is canceled upon summing over the saddle points. We stress that, in this case, the ratio of two `t Hooft coupling goes to zero, equivalently, infinite. The limit decouples dynamics of the two quiver gauge groups and render the global gauge symmetry as a newly emergent flavor symmetry. The non-exponential behavior of the Wilson loop originates from the decoupling, as can be understood intuitively from the screening phenomenon.

\item In $\hat{A}_1$ quiver gauge theory, the two Wilson loops $\langle W_1 \rangle, \langle W_2 \rangle$ associated with the two quiver nodes exhibit the same exponential growth as the ${\cal N}=4$ counterpart. The exponents depend not only on the largest edge of the eigenvalue distribution but also on the two `t Hooft coupling constants, $\lambda_1, \lambda_2$, equivalently, $\lambda, B$.

\item In $\hat{A}_1$ quiver gauge theory, in case the two `t Hooft couplings are the same, so are the two Wilson loops. If the two `t Hooft couplings differ {\sl but} remain finite, the two Wilson loops will also differ. As such, $\langle W_1 \rangle - \langle W_2 \rangle$ is an order parameter of the $\mathbb{Z}_2$ parity exchanging the two quiver nodes. It scales linearly with $B$ and shows {\sl non-analyticity} over the fundamental domain $[-{1 \over 2}, + {1 \over 2}]$.

\end{list}

In this section, we pose these features from holographic dual viewpoint and extract several new perspectives. Much of success of the AdS/CFT correspondence was based on the observation that holographic dual geometry is macroscopically large compared to the string scale. In this limit, string scale effects are suppressed and physical observables and correlators are computable in saddle-point, supergravity approximation. For example, the AdS$_5 \times \mathbb{S}^5$ dual to the ${\cal N}=4$ super Yang-Mills theory has the size $R^2 = O(\sqrt{\lambda})$:
\bea
\rmd s^2 = R^2 \rmd s^2 (\mbox{AdS}_5) + R^2 \rmd \Omega_5^2 (\mathbb{S}^5),
\eea
growing arbitrarily large at strong `t Hooft coupling. Many other examples of the AdS/CFT correspondence share essentially the same behavior. In such a background, expectation value of the Wilson loop $\langle W \rangle$ is evaluated by the Polyakov path integral of a fundamental string in the holographic dual background:
\bea
\langle W \rangle := \int_C [{\cal D} X {\cal D} h]^\perp \, \exp ( i S_{\rm ws}[X^*g])
\label{polyakov}
\eea
with a prescribed boundary condition along the contour $C$ of the Wilson loop at timelike infinity. The worldsheet coupling parameter is set by the pull-back of the spacetime metric, and hence by $R^2$. As $R$ grows large at strong `t Hooft coupling, the path integral is dominated by a saddle point and $\langle W \rangle$ exhibits exponential growth whose Euclidean geometry is the minimal surface ${\cal A}_{\rm cl}$:
\bea
\langle W \rangle \ \ \simeq \ \ e^{{\cal A}_{\rm cl}} \qquad \mbox{where} \qquad
{\cal A}_{\rm cl} \simeq O(R^2) \ .
\eea
Note that the minimal surface of the Wilson loop sweeps out an AdS$_3$ foliation inside the AdS$_5$. This explains the $R^2$ growth of the area of the minimal surface at strong `t Hooft coupling.

Central to our discussions will consist of re-examination on global geometry of the gravity dual to ${\cal N}=2$ superconformal gauge theories in comparison to ${\cal N}=4$ super Yang-Mills theory.

%%%%%%%%%%%%%%%%%%%%%%%%%%%%%%%%%%%%%%%%%%%%%%%%%%%%%%%%%%%%%%%%%%%%%%%%%%%%%%%%%%%%%%%%%%%%%%%%%%%%%%%%%%
\subsection{Holographic dual of $A_1$ gauge theory}
At present, gravity dual to the $A_1$ gauge theory is not known. Still, it is not difficult to guess what the
dual theory would be. In general, ${\cal N}=2$ gauge theory is defined in perturbation theory by three
coupling parameters:
\bea
\lambda, \qquad g_{\rm c}^2 :=  {1 \over N^2}, \qquad g_{\rm o} := {N_{\rm f} \over N},
\eea
associated `t Hooft coupling, closed surface coupling associated with adjoint vector and hypermultiplets, and open puncture coupling associated with fundamental hypermultiplets. For $A_1$ gauge theory,
$g_{\rm o} = 2 \sim O(1)$ and it indicates that dual string theory is described by the worldsheet with proliferating open boundaries. Moreover, as we studied in earlier sections, the $A_1$ gauge theory is related to the $\hat{A}_1$ quiver gauge theory as the limit where one of the two `t Hooft coupling constants is sent to zero while the other is held finite. Equivalently, in the large $N$ limit, one of the two `t Hooft coupling constants is dialed infinitely stronger than the other. This hierarchical scaling limit of the two `t Hooft coupling constants, along with the PSU$(2, 2 \vert 2)$ superconformal symmetry and the SU(2)$\times$U(1) R-symmetry imply that the gravity dual is a noncritical superstring theory involving AdS$_5$ and $\mathbb{S}^2 \times \mathbb{S}^1$ space.
One thus expects that the gravity dual of $A_1$ gauge theory has the local geometry of the form:
\bea
(\mbox{AdS}_5 \times {\cal M}_2) \times [\mathbb{S}^1 \times \mathbb{S}^2] \ .
\label{a1dual}
\eea
By local geometry, we mean that the internal space consists of $\mathbb{S}^1$ and $\mathbb{S}^2$, possibly fibered or warped over an appropriate 2-dimensional base-space ${\cal M}_2$ \footnote{The expected gravity dual (\ref{a1dual}) may be anticipated from the Argyres-Seiberg S-duality \cite{ASduality}. At finite $N$, S-duality maps an infinite coupling ${\cal N}=2$ superconformal gauge theory to a weak coupling ${\cal N}=2$ gauge theory combined with strongly interacting, isolated conformal field theory. The presence of the strongly interacting, isolated conformal field theory suggests that putative holographic dual ought to involve a string geometry whose size is typically of order $O(1)$ in string unit.}. The curvature scales of AdS$_5$ and of ${\cal M}_2$ are equal and are set by $R \sim \lambda^{1/4}$, much as in the ${\cal N}=4$ super Yang-Mills theory. The remaining internal geometry $[\mathbb{S}^1 \times \mathbb{S}^2]$ involves geometry of string scale, and is describable in terms of a (singular) superconformal field theory. In particular, the internal space $[\mathbb{S}^1 \times \mathbb{S}^2]$ may have collapsed 2-cycles.
Therefore, the ten-dimensional geometry is schematically given by
\bea
\rmd s^2 = R^2 (\rmd s^2 (\mbox{AdS}_5) + \rmd s^2 ({\cal M}_2)) + r^2 \rmd s^2 ([\mathbb{S}^1 \times \mathbb{S}^2])
\eea
where $R, r$ are the curvature radii that are hierarchically different, $r \ll R$ (measured in string scale). In particular, $r$ can become smaller than ${\cal O}(1)$ in the regime that the two `t Hooft coupling constants are taken hierarchically disparate.

Consider now evaluating the Wilson loop $\langle W[C] \rangle$ in the gravity dual (\ref{a1dual}). As well-known, the Wilson loop is holographically computed by free energy of a macroscopic string whose endpoint sweeps the contour $C$. From the viewpoint of evaluating it in terms of a minimal area worldsheet, since the internal space has nontrivial 2-cycles, there will not be just one saddle-point but infinitely many. These saddle-point configurations are approximately a combination of minimal surface of area ${\cal A}_{\rm sw}$ inside the AdS$_5$ and surfaces of area $a_{\rm sw}^{(i)}$ wrapping 2-cycles inside the internal space multiple times.
Note that ${\cal A}_{\rm sw}$ has the area of order $O(r^2) \gg 1$ in string unit and $a_{\rm sw}^{(i)}$ has the area of order $O(1)$ since the 2-cycles are collapsed. Therefore, all these configurations have nearly degenerate total worldsheet area and correspond to infinitely many, nearby saddle points. In effect, the surfaces of area $a_{\rm sw}^{(i)}$
wrapping the collapsed 2-cycle multiple times produce sizable worldsheet instanton effects. We thus have
\bea
\langle W \rangle &=& \sum_{i = \rm saddles} c_a \, \exp \left( {\cal A}_{\rm sw} + a_{\rm sw}^{(i)} + \cdots \right)  \nonumber \\
&\simeq& \Big[ \sum_{i = \rm saddles} c_a \, \exp ( a_{\rm sw}^{(i)} ) \Big] \cdot \exp \left( {\cal A}_{\rm sw} \right), \label{summingup}
\eea
where $c_a$ denotes calculable coefficients of each saddle-point, including one-loop string worldsheet determinants and integrals over moduli parameters, if present. This is depicted in figure 6. Since we do not have exact worldsheet result for each saddle point configurations available, we can only guess what must happen in order for the final result to yield the exact result we derived from the gauge theory side. In the last expression of (\ref{summingup}), even though contribution of individual saddle point is same order, summing up infinitely many of them could produce an exponentially small effect of order $O(\exp (-{\cal A}_{\rm sw}))$. What then happens is that summing up infinitely many worldsheet instantons over the internal space cancels against the leading $O(\exp ( {\cal A}_{\rm sw}))$ contribution from the worldsheet inside the AdS$_5$. After the cancelation, the leading nonzero contribution is of the same order as the pre-exponential contribution. It scales as $R^\nu$ for some {\sl finite} value of the exponent $\nu$ at strong `t Hooft coupling.

\begin{figure}[ht!]
\vskip1cm
\centering
\includegraphics[scale=0.68]{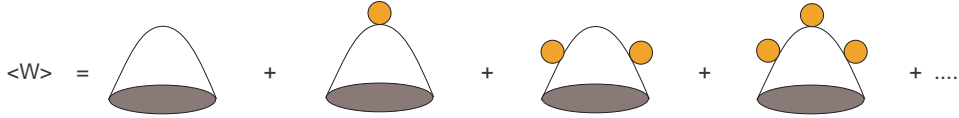}
\caption{\small \sl
Schematic view of holographic computation of Wilson loop expectation value in instanton expansion. Each hemisphere represents minimal surface of semiclassical string in AdS spacetime. Instantons are  string worldsheets $\mathbb{P}^1$'s stretched into the internal space $X_5$. Their sizes are of string scale, and hence of order ${\cal O}(1)$ for any number of instantons. The gauge theory computations indicate that these worldsheet instantons ought to proliferate and lead to delicate cancelations of the leading-order result (the first term) upon resummation.}
\label{}
\vskip1cm
\end{figure}
%

%As noted above, our assertion is that the gravity dual to the $A_1$ gauge theory must involve string-scale geometry and collapsed 2-cycles. While direct construction of Type IIB background is not available yet, a recent study in terms of wrapped M5-brane hints that the gravity dual indeed contains string-scale geometries. In the construction of \cite{gaiottomaldacena}, the M-theory dual to $A_1$ quiver gauge theories is given by AdS$_5 \times {\cal M}_6$ where ${\cal M}_6$ is an $\mathbb{S}^4$ fibered over the Riemann surface $\Sigma_g$ that the M5-branes wrap on. In particular, the internal geometry (\ref{a1dual}) has $A_{2N-1}$ singularities (representing $2N$ fundamental hypermultiplets). In the limit $N \rightarrow \infty$, they provide for infinitely many, vanishing 2-cycles $C_a, (a=1,2, \cdots)$ and in turn infinitely many saddle-points of minimal surfaces by wrapping the string worldsheet on these 2-cycles.

At the orbifold fixed point, there are in general torsion components of the NS-NS 2-form potential $B_2$, whose integral over a 2-cycle is denoted by $B$:
\bea
B_a := \oint_{C_a} {B_2 \over 2 \pi}, \qquad B_a = [0, 1)
\eea
The $A_1$ theory has the global flavor symmetry $G_{\rm f} =$ U$(N_{\rm f}) =$ U$(2N)$.
For a well-defined conformal field theory of the internal geometry, $B_a$ must take the value $1/2$.
But then, the string worldsheet wrapping the 2-cycle $C_a$ $n_a$ times picks up the phase factor
\bea
\prod_{a=1}^\infty \exp (2 \pi i B_a n_a) = \prod_{a=1}^\infty (-)^{n_a},
\eea
giving rise to $\pm$ relative signs among various worldsheet instanton contributions to the minimal surface dual to the Wilson loop.

%%%%%%%%%%%%%%%%%%%%%%%%%%%%%%%%%%%%%%%%%%%%%%%%%%%%%%%%%%%%%%%%%%%%%%%%%%%%%%%%%%%%%%%%%%%%%%%%%%%%%%%%%%%
\subsection{Holographic dual of $\hat{A}_1$ quiver gauge theory}
Consider next holographic description of the $\hat{A}_1$ quiver gauge theory. It is known that the holographic dual is provided by the AdS$_5 \times \mathbb{S}^5/\mathbb{Z}_2$ orbifold, where the $\mathbb{Z}_2$ acts on $\mathbb{C}^2 \subset \mathbb{C}^3$ of the covering space of $\mathbb{S}^5$. Locally, the spacetime geometry is exactly the same as AdS$_5 \times \mathbb{S}^5$:
\bea
\rmd s^2 = R^2 \rmd s^2 (\mbox{AdS}_5) + R^2 \rmd \Omega_5^2 (\mathbb{S}^5).
\eea
The size of both the $AdS_5$ and the $\mathbb{S}^5/\mathbb{Z}_2$ is $R$, which grows as $(\lambda)^{1/4}$ at large `t Hooft coupling limit.

Located at the orbifold fixed point is a twisted sector. The massless fields of the twisted sector consists of a tensor multiplet of $(5+1)$-dimensional (2,0) chiral supersymmetry. The multiplet contains five massless scalars. Three of them are associated with $\mathbb{S}^2$ replacing the orbifold fixed point, and the other two are associated with %
\bea
B = \oint_{\mathbb{S}^2} {B_2 \over 2 \pi} \qquad \mbox{and} \qquad
C = \oint_{\mathbb{S}^2} {C_2 \over 2 \pi},
\eea
where $B_2, C_2$ are NS-NS and R-R 2-form potentials. Both of them are periodic, ranging over $B, C = [0, 1)$ \footnote{The periodicity can be seen from the T-dual, brane configuration as well. Consider the moduli $B$. The quiver gauge theories are mapped to D4 branes connecting adjacent NS5 branes on a circle in two different directions. The sum over gauge couplings is then related to circle size,
while the difference between adjacent gauge couplings is given by the length of each interval. Evidently,
the interval cannot be longer than the circumference.}.
These two massless moduli are well-defined even in the limit that the other three moduli vanish, viz. $\mathbb{S}^2$ shrinks back to the orbifold singularity.
Along with the type IIB dilaton and axion of the untwisted sector, these two twisted scalar
fields are related to the gauge theory parameters. In particular, we have
\bea
{1 \over g_s} = {1 \over g_1^2} + {1 \over g_2^2}; \qquad
{1 \over g_s} (B - {1 \over 2}) = {1 \over g_1^2} - {1 \over g_2^2}.
\label{relation}
\eea
The other moduli field $C$ is related to the theta angles. This can be seen by uplifting the brane configuration to M-theory. There, the theta angle is nothing but the M-theory circle. It would vary if we turn on C-potential on two cycles.

Consider now computation of the Wilson loop expectation value from the Polyakov path integral (\ref{polyakov}). Again, as the contour $C$ of the Wilson loop lies at the boundary of AdS$_3$ foliation inside AdS$_5$, the Type IIB string worldsheet would sweep a minimal surface in AdS$_3$. The area is of order $O(R^2)$. On the other hand,
the Type IIB string may sweep over the vanishing $\mathbb{S}^2$ at the orbifold fixed point. As the area of the cycle vanishes, the corresponding worldsheet instanton effect is of order $O(1)$ and unsuppressed. Thus, the situation is similar to the $A_1$ case. In the $\hat{A}_1$ case, however, we have a new direction of turning on the twisted moduli associated with $B$. From (\ref{relation}), we see that this amounts to turning on the two gauge couplings asymmetrically. Now, for the worldsheet instanton configuration, the Type IIB string worldsheet couples to the $B_2$ field. Therefore, the Wilson loop will get contributions of $\exp ( \pm 2 \pi i B)$ once the moduli $B$ is turned on.

There is another reason why infinitely many worldsheet instantons needs to be resummed. We proved that the
twisted sector Wilson loop is proportional to $| B |$. As $B$ ranges over the interval $[-{1 \over 2}, + {1 \over 2}]$, we see that the Wilson loop has nonanalytic behavior at $B = 0$. In gravity
dual, we argued that the Wilson loop depends on $B$ through the string worldsheet sweeping vanishing two-cycle at the orbifold fixed point. The $n$ instanton effect is proportional to $\exp (2 \pi i n B)$ for $n = \pm 1, \pm 2, \cdots$. It shows that $B$ has the periodicity over $[-{1 \over 2}, +{1 \over 2}]$ and effect of individual instanton is analytic over the period. Obviously, in order to exhibit non-analyticity
such as $|B|$, infinitely many instanton effects needs to be resummed.

%%%%%%%%%%%%%%%%%%%%%%%%%%%%%%%%%%%%%%%%%%%%%%%%%%%%%%%%%%%%%%%%%%%%%%%%%%%%%%%%%%%%%%%%%%%%%%%%%%%%%%%%%%%%
\subsection{Comments on Wilson loops in Higgs phase}
Starting from the $\hat{A}_1$ quiver gauge theory, we have another limit we can take. Consider now the
D3-branes displaced away from the orbifold singularity. If all the branes are moved to a smooth point,
then the quiver gauge symmetry $G$ is broken to the diagonal subgroup $G_{\rm D}$:
\bea
G = {\rm U}(N) \times {\rm U}(N) \qquad \rightarrow  \qquad G_{\rm D} = {\rm U}_{\rm D}(N)
\eea
modulo center-of-mass U(1) group. Of the two bifundamental hypermultiplets, one of them is Higgsed away and the other forms a hypermultiplet transforming in adjoint representation of the diagonal subgroup. This theory flows in the infrared below the Higgs scale to the ${\cal N}=4$ superconformal Yang-Mills theory, as expected since the $N$ D3-branes are stacked now at a smooth point.

We should be able to understand the two Wilson loops of the $\hat{A}_1$ quiver gauge theory in this limit. Obviously, the two Wilson loops $W_1, W_2$ are independent and distinguishable at an energy above the Higgs scale, while they are reduced to one and the same Wilson loop at an energy below the Higgs scale. Noting that Higgs scale is set by the location of the D3-branes from the orbifold singularity, we therefore see that the minimal surface of the macroscopic string worldsheet must exhibit a crossover. How this crossover takes place is a very interesting problem left for the future.

The above consideration is also generalizable to various partial breaking patterns such as
\bea
{\rm SU}(2N) \times {\rm SU}(2N) \rightarrow {\rm SU}(N) \times {\rm SU}(N) \times {\rm SU}_{\rm D}(N) \ .
\eea
Now, there are several types of strings. There are strings corresponding to Wilson loops of three SU($N$)'s. There are also W-bosons that connect diagonal SU($N$) to either of the two SU($N$)'s. The fields now transform as $({\bf N}, \overline{\bf N}; {\bf 1}), (\overline{\bf N}, {\bf N}; {\bf 1})$ and $({\bf 1}, {\bf 1}, {\bf N}^2 - 1)$. As the theory is Higgsed, localization method we relied on is no longer valid. Still, Nevertheless, taking holographic geometry of the conformal points of quiver gauge theories as the starting point, the gravity dual is expected to be a certain class of multi-centered deformations. We expect that one can still learn a lot of (quiver) gauge theory dynamics by taking suitable approximate gravity duals and then computing Wilson loop expectation values and comparing them with weak `t Hooft coupling perturbative results.

\vspace{1cm}
%%%%%%%%%%%%%%%%%%%%%%%%%%%%%%%%%%%%%%%%%%%%%%%%%%%%%%%%%%%%%%%%%%%%%%%%%%%%%%%%%%%%%%%%%%%%%%%%%%%%%%%%%%%%
\section{Generalization to $\hat{A}_{k-1}$ Quiver Gauge Theories}
So far, we were mainly concerned with $A_1$ and $\hat{A}_1$ of ${\cal N}=2$ (quiver) gauge theories.
These are the simplest two within a series of $\hat{A}_{k-1}$ type.
These quiver gauge theories are obtainable from D3-branes sitting at the orbifold singularity $\mathbb{C} \times (\mathbb{C}^2/\mathbb{Z}_k)$. There are $(k-1)$ orbifold fixed points whose blow-up consists of $\mathbb{S}^2_i$ $(i=1, \cdots, k-1)$.
The twisted sector of the Type IIB string theory includes $(k-1)$ tensor multiplets of $(5+1)$-dimensional (2,0) chiral supersymmetry.
Two sets of $(k-1)$ scalar fields are associated with
\bea
B_i = \oint_{\mathbb{S}^2_i} {B_2 \over 2 \pi} \qquad \mbox{and} \qquad
C_i = \oint_{\mathbb{S}^2_i} {C_2 \over 2 \pi} \qquad (i = 1, \cdots, k-1).
\label{2moduli}
\eea
Again, after T-duality to Type IIA string theory, we obtain the $\hat{A}_{k-1}$ brane configuration. As for $k=2$, we first partially compactify the orbifold to $\mathbb{S}^1$ of a fixed asymptotic radius and resolve the $\hat{A}_{k-1}$ singularities. This results in a hyperk\"ahler space where the $\mathbb{S}^1$ is fibered over the base space $\mathbb{R}^3$. The manifold is known as $k$-centered Taub-NUT space. There are $3(k-1)$ geometric moduli associated with $(k-1)$ degeneration centers (where the $\mathbb{S}^1$ fiber degenerates) which, along with the $2(k-1)$ moduli in (\ref{2moduli}), constitute 5 scalar fields of the aforementioned $(k-1)$ tensor multiplets. Now, T-dualizing along the $\mathbb{S}^1$ fiber, we obtain Type IIA background involving $k$ NS5-branes, which source nontrivial dilaton and NS-NS $H_3$ field strength, sitting at the degeneration centers on the base space $\mathbb{R}^3$ and at various positions on the T-dual circle $\widetilde{\mathbb{S}}^1$ set by the $B_i$'s in (\ref{2moduli}).

In the Type IIA brane configuration, there are various limits where global symmetries are enhanced. At generic distribution of $k$ NS5-branes on the dual circle $\widehat{\mathbb{S}}^1$, the global symmetry is given by SU$(2) \times $U(1) associated with the base space $\mathbb{R}^3$ and the dual circle $\widehat{\mathbb{S}}^1$. When (fraction of) NS5-branes all coalesce together, the space transverse to the NS5-branes approaches $\mathbb{C}^2$ very close to them and the U$(1)$ symmetry is enhanced to SU(2). In this limit, (a subset of) gauge couplings of D4-branes become zero and we have global symmetry enhancement. It is well known that $k$-stack of NS5-branes, which source the dilation and the NS-NS $H_3$ field strength, generate the near-horizon geometry of linear dilaton \cite{lineardilaton}.
In string frame, the geometry is the exact conformal field theory \cite{exactCFT}
\bea
\mathbb{R}^{5,1} \times \Big(\mathbb{R}_{\phi, Q} \times {\rm SU}(2)_{k} \Big) \qquad \mbox{where} \qquad Q = \sqrt{2 \over k} \ .
\label{liouville}
\eea
Modulo the center of mass part, the worldvolume dynamics on D4-branes stretched between various NS5-branes can be described in terms of various boundary states \cite{kutasov}, representing localized and extended states in the bulk.

The string theory in this background breaks down at the location of NS5-branes, as the string coupling becomes infinitely strong. To regularize the geometry and define the string theory, we may take $\mathbb{C}$ inside the aforementioned near-horizon $\mathbb{C}^2$, split the coincident $k$ NS5-branes at the center and array them on a concentric circle of a nonzero radius. The string coupling is then cut off at a value set by the radius. The resulting worldsheet theory is the ${\cal N}=2$ supersymmetric Liouville theory.

In the regime we are interested in, $k$ takes values larger than $2$, $k =3, 4, \cdots$. In this regime, the ${\cal N}=2$ Liouville theory (\ref{liouville}) is strongly coupled. By the supersymmetric extension of the Fateev-Zamolodchikov-Zamolodchikov (FZZ) duality, we can turn the ${\cal N}=2$ supersymmetric Liouville theory to Kazama-Suzuki coset theory. To do so, we T-dualize along the angular direction of the arrayed NS5-branes. Conserved winding modes around the angular direction is mapped to
conserved momentum modes and the resulting Type IIB background is given by another
exact conformal field theory
\bea
\mathbb{R}^{5,1} \times \Big({{\rm SL}(2; \mathbb{R})_k \over {\rm U}(1)} \times {{\rm SU}(2)_k \over {\rm U}(1)} \Big)
\label{FZZdual}
\eea
modulo $\mathbb{Z}_k$ orbifolding. For large $k$, the conformal field theory is weakly coupled and describes the well-known cigar geometry \cite{GK2}.

In the large (finite or infinite) $k$, what do we expect for the Wilson loop expectation value and, from the expectation values, what information can we extract for the holographic geometry of gravity dual? Here, we
shall remark several essential points that are extendible straightforwardly from the results of $\hat{A}_1$
and relegate further aspects in a separate work. For $\hat{A}_{k-1}$ quiver gauge theories,
there are $k$ nodes of gauge groups U($N$). Associated with them are $k$ independent Wilson loops:
\bea
W^{(i)}[C] := \mbox{Tr}_{(i)}P_s  \exp\Bigl[ ig \int_C \rmd \left(\dot{x}^mA^{(i)}_m (x) +\theta^a A_a^{(i)}(x) \right) \Bigr] \qquad (i=1, \cdots, k) \ .
\eea
From these, we can construct the Wilson loop in untwisted and twisted sectors. Explicitly, they are
\bea
W_0 = {1 \over k} \Big( W^{(1)}+ W^{(2)}+ \cdots + W^{(k-1)} + W^{(k)} \Big)
\eea
for the untwisted sector Wilson loop and
\bea
&& W_1 = W^{(1)} + \omega W^{(2)} + \cdots + \omega^{k-1} W^{(k)} \nonumber \\
&& W_2 = W^{(1)} + \omega^2 W^{(2)} + \cdots + \omega^{2(k-1)} W^{(k)} \nonumber \\
&& \cdots \nonumber \\
&& W_{k-1} = W^{(1)} + \omega^{k-1} W^{(2)} + \cdots + \omega^{(k-1)^2} W^{(k)}
\eea
for the $(k-1)$ independent twisted sector Wilson loops. They are simply $k$ normal modes of Wilson loops constructed from $\{ \omega^n \vert n=0, \cdots, k-1\}$ Fourier series of $\mathbb{Z}_k$ over the $k$ quiver nodes.
Consider now the planar limit $N \rightarrow \infty$. The Wilson loops $W^{(i)}$ are all same. Equivalently, all the twisted Wilson loops vanish. Furthermore, as in $\hat{A}_1$ quiver gauge theory, the untwisted Wilson loop will show exponential growth at large `t Hooft coupling.

It is not difficult to extend the gauge theory results to $\hat{A}_{k-1}$ case. After taking large $N$ limit, the saddle point equations now read
\begin{eqnarray}
\frac{4\pi^2}{\lambda}\phi &=& \int\hspace{-4mm}-\hspace{2.5mm}\rmd\phi'\frac{\rho(\phi')}{\phi-\phi'},
   \label{A_2} \label{akuntwisted} \\
\frac{2\pi^2}{\lambda_a} \phi
 -(1-\overline{\omega})\int\hspace{-4mm}-\hspace{2.5mm}\rmd\phi'\delta_a\rho(\phi')F'(\phi-\phi')
 &=& \int\hspace{-4mm}-\hspace{2.5mm}\rmd\phi'\frac{\delta_a \rho(\phi')}{\phi-\phi'},
 \qquad (a=1, \cdots, k-1) \label{aktwisted} \nonumber \\
\end{eqnarray}
where
\bea
\rho &:=& {1 \over k} \Big( \rho^{(1)} + \cdots + \rho^{(k)} \Big) \nonumber \\
\delta_a \rho &:=& {1 \over k} \sum_{i=1}^k \omega^{i-1} \rho^{(i)} \qquad
(a=1,2, \cdots, k-1),
\eea
and
\bea
&& {1 \over \lambda} := {1 \over k} \Big( {1 \over \lambda^{(1)}} + \cdots +{1 \over \lambda^{(k)}} \Big) \nonumber \\
&& {1 \over \lambda}_a := {1 \over k} \sum_{i=1}^{k} \omega^{i-1} {1 \over \lambda^{(i)}}  \qquad
(a = 1, 2, \cdots, k-1).
\eea
It is evident that $\delta_a \rho$ is proportional to $1/\lambda_a$ linearly, and hence exhibits {\sl non-analytic} behavior.

By the AdS/CFT correspondence, the Wilson loops are mapped to macroscopic fundamental Type IIB string in
the geometry AdS$_5 \times \mathbb{S}^5/\mathbb{Z}_k$. There are $(k-1)$ 2-cycles of vanishing volume.
As in the $\hat{A}_1$ case, $n$ worldsheet instanton picks up a phase factor exp$(2 \pi i B n)$. Again, since $B=1/2$ for the exact conformal field theory, the phase factor is given by $(-)^n$. As (fraction of) the gauge couplings are tuned to zero, we again see from (\ref{aktwisted}) that twisted Wilson loops are
suppressed by the worldsheet instanton effects. This is the effect of the screening we explained in the previous section, but now extended to the $\hat{A}_{k-1}$ quiver theories. The suppression, however, is less significant as $k$ becomes large since the one-loop contribution in (\ref{aktwisted}) is hierarchically small compared to the classical contribution. We see this as a manifestation of the fact we recalled above that, at $k \rightarrow \infty$, the worldsheet conformal field theory is weakly coupled in Type IIB setup and the holographic dual geometry, the cigar geometry, becomes weakly curved.

It is also illuminating to understand the above Wilson loops from the viewpoint of the brane configuration.
For the brane configuration,
 we start from the Type IIA theory on a compact spatial circle of circumference $L$.
We place $k$ NS5-branes on the circle on intervals $L_a, (a=1, 2, \cdots, k)$ such that $L_1 + L_2 + \cdots + L_k = L$ and then stretch $N$ D4-branes on each interval. The low-energy dynamics of these D4-branes is then described by ${\cal N}=2$ quiver gauge theory of $\hat{A}_{k-1}$ type. In this setup, the $W^{(a)}$ Wilson loop is represented by a semi-infinite, macroscopic string emanating from $a$-th D4-brane to infinity. Since there are $k$ different states for identical macroscopic strings, we can also form linear combinations of them. There are $k$ different normal modes: the untwisted Wilson loop $W_0$ is the lowest normal mode obtained by algebraic average of the $k$ strings, $W_1$ is the next lowest normal mode obtained by discrete lattice translation $\omega$ for adjacent strings, $\cdots$, and the $W_{k-1}$ is the highest normal mode obtained by discrete lattice translation $\omega^{k-1}$ (which is the same as the configuration with lattice momentum $\bar{\omega}$ by the Unklapp process) for adjacent strings.

If the intervals are all equal, $L_1 = L_2 = \cdots = L_k = (L/k)$, then the brane configuration has cyclic permutation symmetry. This symmetry then ensures that all twisted Wilson loops vanish. If the intervals are different, (some of) the twisted Wilson loops are non-vanishing. If (fraction of) NS5-branes become coalescing, the geometry and the worldvolume global symmetries get enhanced. We see that fundamental strings ending on the weakly coupled D4-branes will be pulled to the coalescing NS5-branes. The difference from the $A_1$ theory is that, effect of other NS5-branes away from the coalescing ones becomes larger as $k$ gets larger. This is the brane configuration counterpart of the suppression of twisted Wilson loop expectation value which were attributed earlier to the weak curvature of the holographic geometry (\ref{FZZdual}) in this limit.

%%%%%%%%%%%%%%%%%%%%%%%%%%%%%%%%%%%%%%%%%%%%%%%%%%%%%%%%%%%%%%%%%%%%%%%%%%%%%%%%%%%%%%%%%%%%%%%%%%%%%%%%%%%
\section{Discussion} \label{discuss}

\vspace{5mm}
In this paper, we investigated aspects of four-dimensional ${\cal N}=2$ superconformal gauge theories.
Utilizing the localization technique, we showed that the path integral of these theories are reduced to a finite-dimensional matrix integral, much as for the ${\cal N}=4$ super Yang-Mills theory. The resulting matrix model is, however, non-Gaussian. Expectation value of half-BPS Wilson loops in these theories can also be evaluated using the matrix model techniques. We studied two theories in detail:
$A_1$ gauge theory with gauge group U$(N)$ and $2N$ fundamental hypermultiplets and $\hat{A}_1$ quiver gauge theory with gauge group U$(N)\times$U$(N)$ and two bi-fundamental hypermultiplets.

In the planar limit, $N \rightarrow \infty$, we determined exactly the leading asymptotes of the circular Wilson loops as the `t Hooft coupling becomes strong, $\lambda \rightarrow \infty$ and then compared it to the exponential growth $\sim \exp(\sqrt{\lambda})$ seen in the ${\cal N}=4$ super Yang-Mills theory.
In the $A_1$ theory, we found the Wilson loop exhibits {\sl non-exponential} growth:
it is bounded from above in the large $\lambda$ limit.
In the $\hat{A}_1$ theory, there are two Wilson loops, corresponding to the two $U(N)$ gauge groups. We found that the untwisted Wilson loop exhibits exponential growth, exactly the same leading behavior as the Wilson loop in ${\cal N}=4$ super Yang-Mills theory, but the twisted Wilson loop exhibits a new {\sl non-analytic} behavior in difference of the two gauge coupling constants.

We also studied holographic dual of these ${\cal N}=2$ theories and macroscopic string configurations representing the Wilson loops. We argued that both the {\sl non-exponential} behavior of the $A_1$ Wilson loop and the {\sl non-analytic} behavior of the $\hat{A}_1$ Wilson loops are indicative of string scale geometries of the gravity dual. For gravity dual of $A_1$ theory, there are infinitely many vanishing 2-cycles around which the macroscopic string wraps around and produce worldsheet instantons. These different saddle-points interfere among themselves, canceling out the would-be leading exponential growth. What remains thereafter then yields a non-exponential behavior, matching with the exact gauge theory results. For gravity dual of $\hat{A}_1$ theory, there is again a  vanishing 2-cycle at the $\mathbb{Z}_2$ orbifold singularity. On the 2-cycle, NS-NS 2-form potential can be turned on and it is set by asymmetry between the two gauge coupling constants. The macroscopic string wraps around and each worldsheet instanton is weighted by $\exp (2 \pi i B)$. Again, since the 2-cycle has a vanishing area, infinite number of worldsheet instantons needs to be resummed. The resummation can then yield a non-analytic dependence on $B$, and this fits well with the exact gauge theory result.

A key lesson drawn from the present work is that holographic dual of these ${\cal N}=2$ superconformal gauge theories must involve geometry of string scale. For $A_1$ theory, suppression of exponential growth of Wilson loop expectation value hints that the holographic duals must be a noncritical string theory. In the brane construction viewpoint, this arose because the two coinciding NS5-branes generates the well-known linear dilaton background near the horizon and macroscopic string is pulled to the NS5-branes. In the holographic dual gravity viewpoint, this arose because worldsheet of macroscopic string representing the Wilson loop is not peaked to a semiclassical saddle-point but is affected by proliferating worldsheet instantons. We argued that delicate cancelation among the instanton sums lead to non-exponential behavior
of the Wilson loop.

It should be possible to extend the analysis in this paper to general ${\cal N}=2$ superconformal gauge theories. Recently, various quiver constructions were put forward \cite{gaiotto} and some of its gravity duals were studied \cite{gaiottomaldacena}. Main focus of this line of research were on quiver generalization of the Argyres-Seiberg S-duality, which does not commute with the large $N$ limit. Aim of the present work was to characterize behavior of the Wilson loop in large $N$ limit in terms of representation contents of matter fields and, from the results, infer the holographic geometry of gravity duals. We also remarked that our approach is complementary to the researches based on various worldsheet formulations \cite{Kawai:2007ek}\cite{Kawai:2007eg}\cite{Berkovits:2007rj}\cite{Azeyanagi:2008mi}.

Recently, localization in the ${\cal N}=6$ superconformal Chern-Simons theory was obtained and Wilson loops therein was studied in detail \cite{ABJMlocalization}. It should also be possible to extend the analysis to the superconformal (quiver) Chern-Simons theories. In particular, given that these two types of theories are related roughly speaking by partially compactifying on $\mathbb{S}^1$ and flowing into infrared, understanding similarities and differences between quiver gauge theories in (3+1) dimensions and in (2+1) dimensions would be extremely useful for elucidating further relations in gauge and string dynamics.

Finally, it should be possible to extend the analysis in this work to ${\cal N}=1$ superconformal quiver gauge theories and study implications to the Seiberg duality. Candidate non-critical string duals of these gauge theories were proposed by \cite{israel}.

We are currently investigating these issues but will relegate reporting our findings to follow-up publications.

\vspace{2cm}

\section*{Acknowledgments}

We are grateful to Zoltan Bajnok, Dongsu Bak, David Gross and Juan Maldacena for useful discussions on topics related to this work and comments. SJR thanks Kavli Institute for Theoretical Physics for hospitality during this work. TS thanks KEK Theory Group, Institute for Physics and Mathematics of the Universe and Asia-Pacific Center for Theoretical Physics for hospitality during this work. This work was supported in part by the National Science Foundation of Korea Grants 2005-084-C00003, 2009-008-0372, 2010-220-C00003, EU-FP Marie Curie Research \& Training Networks HPRN-CT-2006-035863 (2009-06318) and U.S. Department of Energy Grant DE-FG02-90ER40542.

%%%%%%%%%%%%%%%%%%%%%%%%%%%%%%%%%%%%%%%%%%%%%%%%%%%%%%%%%%%%%%%%%%%%%%%%%%%%%%%%%%%%%%%%%%%%%%%%%%%%%%%%%%%%
\appendix

\makeatletter
\@addtoreset{equation}{section}
\def\theequation{\thesection.\arabic{equation}}
\makeatother

\vspace{1cm}

\section{Killing spinor on $\mathbb{S}^4$} \label{spinor}

\vspace{5mm}
The Killing spinors on $\mathbb{S}^4$ are defined as follows. Let $y^a$ $(a=1,\cdots,5)$ be coordinates of $\mathbb{R}^5$. We embed $\mathbb{S}^4$ into $\mathbb{R}^5$ by the hypersurface
\begin{equation}
(y^a+z^a)^2 = r^2, \qquad z^a = (0,\cdots,0,r).
\end{equation}
Each point on $\mathbb{S}^4$ can be mapped to a point on a four-dimensional hyperplane $\mathbb{R}^4$, $y^5=0$, tangent to the North Pole through
\begin{equation}
y^a = -2z^a +e^\Omega(x^a+2z^a), \qquad e^\Omega = \left( 1+\frac{x^2}{4r^2} \right)^{-1},
\end{equation}
where $x^a=(x^m,x^5=0)$. This describes a projection on $\mathbb{R}^4$ from the South Pole of $\mathbb{S}^4$. Accordingly, the induced metric on $\mathbb{S}^4$ is given by
\begin{eqnarray}
\rmd s^2 &=& h_{mn}\, \rmd x^m\, \rmd x^n \nonumber \\
&=& e^{2\Omega}\delta_{mn}\, \rmd x^m\, \rmd x^n.
\end{eqnarray}
Let $\theta$ be the polar angle measured from the North Pole, viz. the origin of the $\mathbb{R}^4$.
Then, for a fixed $\theta$, the coordinates $x^m$ satisfy
\begin{equation}
\sum_{m=1}^4(x^m)^2 = 4r^2\tan^2\frac\theta2.
\end{equation}
We also denote orthonormal frame coordinates as $x^{\hat{m}}$, ($\hat{m} = \hat{1}, \cdots, \hat{4})$ with vierbein $e^{\hat{m}}_m = \delta^{\hat{m}}_m e^\Omega$.

\vspace{5mm}

It is straightforward to show that the spinors
\begin{eqnarray}
\xi &=& e^{\frac12\Omega}(\xi_s+x^{\hat{m}}\Gamma_{\hat{m}}\xi_c),
   \label{varepsilon} \\
\widetilde{\xi} &=& e^{\frac12\Omega}(\xi_c-\frac1{4r^2}x^{\hat{m}}\Gamma_{\hat{m}}\xi_s),
\end{eqnarray}
where $\xi_s$ and $\xi_c$ are arbitrary constant Majorana-Weyl spinors,
satisfy the conformal Killing spinor equations
\begin{equation}
\nabla_m\xi = \Gamma_m\widetilde{\xi}, \qquad \nabla_m\widetilde{\xi} = -\frac1{4r^2}\Gamma_m\xi \, .
\end{equation}

We further impose anti-chirality condition:
\begin{equation}
\Gamma^{\hat{1}\hat{2}\hat{3}\hat{4}}\xi_s = -\xi_s, \qquad \xi_c = \frac1{2r}\Gamma^{0\hat{1}\hat{2}}\xi_s.
\end{equation}
These equations imply
\begin{equation}
\overline{\xi}\widetilde{\xi} = 0, \qquad \overline{\xi}\Gamma^{05}\widetilde{\xi} = 0.
\end{equation}

One can show that the components of $v^M=\overline{\xi}\Gamma^M\xi$ have the following explicit forms:
\begin{eqnarray}
v^1 = \frac{x_2}r, &\hspace{5mm}& v^2 = -\frac{x_1}r, \\
v^3 = \frac{x_4}r, &\hspace{5mm}& v^4 = -\frac{x_3}r, \\
v^0 = -1, &\hspace{5mm}& v^5 = \cos\theta, \\
v^{6,7,8,9} = 0, &&
\end{eqnarray}
where we normalized $\xi_s$ such that $\overline{\xi}_s\Gamma^0\xi_s=-1$.

The expression (\ref{varepsilon}) can be rewritten as follows:
\begin{equation}
\xi = e^{\frac12\Omega}\xi_s+\frac12e^{-\frac12\Omega}v_{\hat{m}}\Gamma^{\hat{m}}\Gamma^5\xi_s.
\end{equation}
We define
\begin{equation}
n_{\hat{m}} := \frac{v_{\hat{m}}}{\sin\theta}
\end{equation}
so that
\begin{equation}
(n_{\hat{m}}\Gamma^{\hat{m}}\Gamma^5)^2 = -1.
\end{equation}
Then, it is easy to show that the conformal Killing spinor is expressible as
\begin{eqnarray}
\xi (x)
&=& \left( \cos\frac\theta2+\sin\frac\theta2 n_{\hat{m}}(x)\Gamma^{\hat{m}}\Gamma^5 \right)\xi_s \nonumber \\
&=& \exp\left( \frac\theta2n_{\hat{m}}(x) \Gamma^{\hat{m}}\Gamma^5 \right)\xi_s.
\end{eqnarray}

The conformal Killing spinors $\xi$ and $\widetilde{\xi}$ satisfy the following identities:
\begin{eqnarray}
v^m\nabla_m\xi-\frac12(\bar{\xi}\Gamma_{mn}\tilde{\xi})\Gamma^{mn}\xi+\frac12(\bar{\xi}\Gamma_{st}\tilde{\xi})\Gamma^{st}\xi
&=& 0, \\
v^m\nabla_m\tilde{\xi}-\frac12(\bar{\xi}\Gamma_{mn}\tilde{\xi})\Gamma^{mn}\tilde{\xi}
 +\frac12(\bar{\xi}\Gamma_{st}\tilde{\xi})\Gamma^{st}\tilde{\xi}
&=& 0.
\end{eqnarray}

\vspace{1cm}

\section{Spinors for off-shell closure} \label{spinor2}

\vspace{5mm}

We define
\begin{equation}
\nu_0^{\dot{m}} := \Gamma^{\dot{m}}\Gamma^{\hat{1}}\xi_s, \qquad \nu_0^s := \Gamma^s\Gamma^{\hat{1}}\xi_s,
\end{equation}
where $\dot{m}=\hat{2},\hat{3},\hat{4}$.
Let $I=(\dot{m},s)$.
It can be shown that
\begin{eqnarray}
\overline{\xi}_s\Gamma^M\nu_0^I &=& 0, \\
\overline{\nu}_0^I\Gamma^M\nu_0^J &=& \delta^{IJ}\overline{\xi}_s\Gamma^M\xi_s, \\
\frac12v_s^M \Gamma_M &=& \xi_s\overline{\xi}_s+\nu_0^I\overline{\nu}_{0I}
\end{eqnarray}
hold, where $v_s^M=\overline{\xi}_s\Gamma^M\xi_s$.
Since $\xi$ is obtained from $\xi_s$ through a rotation, if we define
\begin{equation}
\nu^I := \exp\Bigl( \frac\theta2n_{\hat{m}}\Gamma^5\Gamma^{\hat{m}} \Bigr)\nu_0^I,
\end{equation}
then the following relations follow:
\begin{eqnarray}
\overline{\xi}\Gamma^M\nu^I &=& 0, \\
\overline{\nu}^I\Gamma^M\nu^J &=& \delta^{IJ}\overline{\xi}\Gamma^M\xi, \\
\frac12v^M \Gamma_M &=& \xi\overline{\xi}+\nu^I\overline{\nu}_{I}
\end{eqnarray}
If the last equation is projected onto the space of $\lambda$, one finds
\begin{equation}
\frac12v^M\Gamma_M = \xi\bar{\xi}+\nu^{\dot{m}}\overline{\nu}_{\dot{m}},
\end{equation}
while in the space of $\psi$, it becomes
\begin{equation}
\frac12v^M\Gamma_M = \nu_\alpha\overline{\nu}^\alpha.
\end{equation}

The spinors satisfy the following identities:
\begin{eqnarray}
v^m\nabla_m\nu^{\dot{k}}-\frac12(\overline{\xi}\Gamma_{mn}\widetilde{\xi})
\Gamma^{mn}\nu^{\dot{k}}+\frac12(\overline{\xi}\Gamma_{st}\widetilde{\xi})
 \Gamma^{st}\nu^{\dot{k}}+(\overline{\nu}^{\dot{k}}\Gamma^m\nabla_m\nu_{\dot{n}})\nu^{\dot{n}} &=& 0, \\
v^m\nabla_m\nu_\alpha-\frac12(\overline{\xi}\Gamma_{mn}\widetilde{\xi})\Gamma^{mn}\nu_\alpha
 -\nu_\beta\overline{\nu}^\beta\Gamma^m\nabla_m\nu^\alpha &=& 0.
\end{eqnarray}

\vspace{5mm}

Due to the above choice of spinors, $\mathfrak{Q}^2$ closes on fields as follows:
\begin{eqnarray}
-i\, \mathfrak{Q}^2 \, A_m \,
&=& v^n\nabla_nA_m+\nabla_mv^nA_n-ig[v^\mu A_\mu,A_m]-\nabla_m(v^\mu A_\mu), \\
-i\, \mathfrak{Q}^2\, A_a  \,
&=& v^m\nabla_mA_a-ig[v^\mu A_\mu,A_a], \\
-i\, \mathfrak{Q}^2 \, q^\alpha\,
&=& v^m\nabla_mq^\alpha-ig(v^\mu A_\mu^A)T_Aq^\alpha+2\overline{\xi}\gamma^\alpha{}_\beta\widetilde{\xi}q^\beta, \\
-i\, \mathfrak{Q}^2 \, q_\alpha \,
&=& v^m\nabla_mq_\alpha+ig(v^\mu A_\mu^A)q_\alpha T_A-2q_\beta\overline{\xi}\gamma^\beta{}_\alpha\widetilde{\xi}, \\
-i\, \, \mathfrak{Q}^2\, \, \lambda \,\,
&=& v^m\nabla_m\lambda-\frac12(\overline{\xi}\Gamma_{mn}\tilde{\xi})\Gamma^{mn}\lambda-ig[v^\mu A_\mu,\lambda]
    +\frac12(\overline{\xi}\Gamma_{st}\widetilde{\xi})\Gamma^{st}\lambda, \\
-i \, \, \mathfrak{Q}^2 \, \, \psi \, \,
&=& v^m\nabla_m\psi-\frac12(\bar{\xi}\Gamma_{mn}\tilde{\xi})\Gamma^{mn}\psi-ig(v^\mu A_\mu^A)T_A\psi, \\
-i \, \, \mathfrak{Q}^2\, \, \bar{\psi} \, \,
&=& v^m\nabla_m\bar{\psi}+\frac12(\bar{\xi}\Gamma_{mn}\tilde{\xi})\bar{\psi}\Gamma^{mn}+ig(v^\mu A_\mu^A)\bar{\psi}T_A, \\
-i \, \mathfrak{Q}^2   K^{\dot{m}}
&=& v^k\nabla_kK^{\dot{m}}-ig[v^\mu A_\mu,K^{\dot{m}}]+\bar{\nu}^{\dot{m}}\Gamma^k\nabla_k\nu_{\dot{n}}K^{\dot{n}}, \\
-i\, \mathfrak{Q}^2  K^\alpha
&=& v^m\nabla_mK^\alpha-ig(v^\mu A_\mu^A)T_AK^\alpha+\bar{\nu}^\alpha\Gamma^m\nabla_m\nu_\beta K^\beta, \\
-i\, \mathfrak{Q}^2  K_\alpha
&=& v^m\nabla_mK_\alpha+ig(v^\mu A_\mu^A)K_\alpha T_A-K_\beta\bar{\nu}^\beta\Gamma^m\nabla_m\nu_\alpha.
\end{eqnarray}

\vspace{1cm}

\section{Asymptotic expansion of Wilson loop} \label{AEestimate}

\vspace{5mm}
In this appendix, we provide details of the asymptotic expansion of the Wilson loop in the large $a$ limit.

We first estimate the following integral:
\begin{equation}
I(\alpha,a) := \int_\delta^\infty \rmd u \ u^\alpha e^{-au},
\end{equation}
where $a,\alpha,\delta>0$.
This satisfies the relation
\begin{equation}
I(\alpha,a) = \frac{\delta^\alpha}ae^{-\delta a}+\frac{\alpha}aI(\alpha-1,a).
\end{equation}
There exists an integer $K$ for which $\alpha-K+1>0$ and $\alpha-K<0$.
Then, repeating integration by parts, $I(\alpha,a)$ can be written as
\begin{equation}
I(\alpha,a) = \sum_{n=0}^{K-1}\frac{\delta^{\alpha-n}}{a^{n+1}}\frac{\Gamma(\alpha+1)}{\Gamma(\alpha+1-n)}e^{-\delta a}
 +\frac1{a^K}\frac{\Gamma(\alpha+1)}{\Gamma(\alpha+1-K)}I(\alpha-K,a).
\end{equation}
$I(\alpha-K,a)$ is estimated as follows:
\begin{equation}
I(\alpha-K,a) \le \delta^{\alpha-K}\int_\delta^\infty \rmd u\ e^{-au} = \frac{\delta^{\alpha-K}}ae^{-\delta a}.
\end{equation}
Therefore, for large $a$, $I(\alpha,a)$ is estimated to be
\begin{equation}
I(\alpha,a) = O(a^{-1}e^{-\delta a}).
\end{equation}

With the above result, we now estimate $W$. With the assumed behavior of rescaled density function $\tilde{\rho}$ in section 3, one can write $e^{-ca}W$ as
\begin{equation}
\int_0^{1-\frac ab} \rmd u\,\tilde{\rho}(1-u)e^{-cau} = \beta\int_0^\delta \rmd u\ u^\alpha e^{-cau}+\int_0^\delta \rmd u\,\chi(u)e^{-cau}
  + \int_\delta^{1-\frac ab} \rmd u\,\tilde{\rho}(1-u)e^{-cau}.
\end{equation}
The first term of the right-hand side is
\begin{eqnarray}
\beta\int_0^\delta \rmd u\ u^\alpha e^{-cau}
&=& \beta\int_0^\infty \rmd u\ u^\alpha e^{-cau} - \beta I(\alpha,ca) \nonumber \\
&=& \beta\Gamma(\alpha+1)(ca)^{-\alpha-1} + O((ca)^{-1}e^{-\delta ca}).
\end{eqnarray}
The second term can be evaluated similarly, and it turns out to be negligible compared to the first term.
The third term is
\begin{equation}
\int_\delta^{1-\frac ab}\rmd u\,\tilde{\rho}(1-u)e^{-cau} \le e^{-\delta ca}\int_\delta^{1-\frac ab} \rmd u\,\tilde{\rho}(u-1) \le e^{-\delta ca}.
\end{equation}
This completes the proof of the proclaimed estimate (\ref{estimate}) in the large $a$ limit.

\vspace{1cm}

\section{Coefficient $c_1$} \label{coeff}

\vspace{5mm}
In this appendix, we elaborate detailed calculation of the coefficient $c_1$ of the leading term in the one-loop determinant. The heat-kernel coefficient $a_2(\Delta)$ is
\begin{equation}
a_2(\Delta) = \frac1{(4\pi)^2}\int_{{\bf S}^4}\rmd^4 x\, \sqrt{h}\,\mbox{tr}_B\Bigl[ -\frac1{4r^2}(3+\cos^2\theta)+\frac16R \Bigr],
\end{equation}
where $\mbox{tr}_B$ is the trace over the indices $\alpha,\beta$.
The second term is canceled by the fermionic contribution.
The first term yields $-\frac5{12}r^2$.

The coefficient $a_2(\Delta_F)$ for the fermions is
\begin{equation}
a_2(\Delta_F) = \frac1{(4\pi)^2}\int_{{\bf S}^4}\rmd^4 x\sqrt{h}\,\mbox{tr}_F\Bigl[ \frac{3\kappa^2}{r^3}+\frac{\kappa^2}4
 (\bar{\xi}\Gamma_{\mu\nu}\tilde{\xi})(\bar{\xi}\Gamma_{\rho\sigma}\tilde{\xi})\Gamma^{\mu\nu}\Gamma^{\rho\sigma}+\frac16R \Bigr],
\end{equation}
where $\mbox{tr}_F$ is the trace over the subspace of the spinor corresponding to $\psi$.
One can show that the first two terms cancel each other.

As the $-\Delta_F$ has the term linear in $m$, $a_4(\Delta_F)$ also contribute to $c_1$.
The relevant part of the coefficient $a_4(\Delta_F)$ is
\begin{equation}
\frac1{(4\pi)^2}\int_{{\bf S}^4}\rmd^4 x\sqrt{h}\,\mbox{tr}_F\Bigl[ \frac12\Bigl( i\kappa\frac mr(\bar{\xi}\Gamma_{\mu\nu}
 \tilde{\xi})\Gamma^{\mu\nu} \Bigr)^2 \Bigr] = -\frac23m^2.
\end{equation}

As a result, it follows that
\begin{equation}
c_1 = -\Bigl( -\frac5{12} \Bigr)-\frac12\Bigl( -\frac23 \Bigr) = \frac34.
\end{equation}


\vspace{1cm}

\begin{thebibliography}{99}

%\cite{Maldacena:1997re}
\bibitem{Maldacena:1997re}
  J.~M.~Maldacena,
{\sl The large N limit of superconformal field theories and supergravity},
  Adv.\ Theor.\ Math.\ Phys.\  {\bf 2}, 231 (1998)
  [Int.\ J.\ Theor.\ Phys.\  {\bf 38}, 1113 (1999)]
  [arXiv:hep-th/9711200].
  %%CITATION = IJTPB,38,1113;%%

%\cite{Rey:1998ik}
\bibitem{Rey:1998ik}
  S.~J.~Rey and J.~T.~Yee,
{\sl Macroscopic strings as heavy quarks in large N gauge theory and  anti-de
  Sitter supergravity},
  Eur.\ Phys.\ J.\  C {\bf 22} (2001) 379
  [arXiv:hep-th/9803001].
  %%CITATION = EPHJA,C22,379;%%
%\cite{Rey:1998bqp}
  S.~J.~Rey, S.~Theisen and J.~T.~Yee,
{\sl Wilson-Polyakov loop at finite temperature in large N gauge theory and
  anti-de Sitter supergravity},
  Nucl.\ Phys.\  B {\bf 527} (1998) 171
  [arXiv:hep-th/9803135].
  %%CITATION = NUPHA,B527,171;%%

%\cite{Maldacena:1998im}
\bibitem{Maldacena:1998im}
  J.~M.~Maldacena,
{\sl Wilson loops in large N field theories},
  Phys.\ Rev.\ Lett.\  {\bf 80} (1998) 4859
  [arXiv:hep-th/9803002].
  %%CITATION = PRLTA,80,4859;%%

%\cite{Erickson:2000af}
\bibitem{Erickson:2000af}
  J.~K.~Erickson, G.~W.~Semenoff and K.~Zarembo,
{\sl Wilson loops in N = 4 supersymmetric Yang-Mills theory},
  Nucl.\ Phys.\  B {\bf 582}, 155 (2000)
  [arXiv:hep-th/0003055].
  %%CITATION = NUPHA,B582,155;%%

%\cite{Drukker:2000rr}
\bibitem{Drukker:2000rr}
  N.~Drukker and D.~J.~Gross,
{\sl An exact prediction of N = 4 SUSYM theory for string theory},
  J.\ Math.\ Phys.\  {\bf 42} (2001) 2896
  [arXiv:hep-th/0010274].
  %%CITATION = JMAPA,42,2896;%%



%\cite{Pestun:2007rz}
\bibitem{Pestun:2007rz}
  V.~Pestun,
{\sl Localization of gauge theory on a four-sphere and supersymmetric Wilson
  loops},
  arXiv:0712.2824 [hep-th].
  %%CITATION = ARXIV:0712.2824;%%

%\cite{Rey:2008bh}
\bibitem{Rey:2008bh}
S.~J.~Rey, T.~Suyama and S.~Yamaguchi,
  {\sl Wilson Loops in Superconformal Chern-Simons Theory and Fundamental Strings
  in Anti-de Sitter Supergravity Dual},
  JHEP {\bf 0903} (2009) 127
  [arXiv:0809.3786 [hep-th]].
  %%CITATION = JHEPA,0903,127;%%

%\cite{Drukker:2008zx}
\bibitem{Drukker:2008zx}
  N.~Drukker, J.~Plefka and D.~Young,
{\sl Wilson loops in 3-dimensional N=6 supersymmetric Chern-Simons Theory and their string theory duals},
  JHEP {\bf 0811}, 019 (2008)
  [arXiv:0809.2787 [hep-th]].
  %%CITATION = JHEPA,0811,019;%%

%\cite{Chen:2008bp}
\bibitem{Chen:2008bp}
  B.~Chen and J.~B.~Wu,
{\sl Supersymmetric Wilson Loops in N=6 Super Chern-Simons-matter theory},
  arXiv:0809.2863 [hep-th].
  %%CITATION = ARXIV:0809.2863;%%

%\cite{Aharony:2008ug}
\bibitem{Aharony:2008ug}
  O.~Aharony, O.~Bergman, D.~L.~Jafferis and J.~Maldacena,
  {\sl N=6 superconformal Chern-Simons-matter theories, M2-branes and their gravity duals},
  JHEP {\bf 0810}, 091 (2008)
  [arXiv:0806.1218 [hep-th]].
  %%CITATION = JHEPA,0810,091;%%

\bibitem{suyama} T. Suyama, talk given at KEK String Advanced Lectures Workshop (April 17, 2009, Tsukuba, Japan)
{\tt http://research.kek.jp/group/www-theory/theory\hskip-0.5cm\_center/SAL/slides/Suyama\_090417.pdf }
.

\bibitem{rey} S.J. Rey, talk given at Strings 2009 Conference (June 25, 2009, Rome, Italy)
{\tt http://strings2009.roma2.infn.it/talks/Rey\_Strings09.PDF }
.

%\cite{Berkovits:1993zz}
\bibitem{Berkovits:1993zz}
  N.~Berkovits,
{\sl A Ten-dimensional superYang-Mills action with off-shell supersymmetry},
  Phys.\ Lett.\  B {\bf 318}, 104 (1993)
  [arXiv:hep-th/9308128].
  %%CITATION = PHLTA,B318,104;%%

%\cite{Evans:1994cb}
\bibitem{Evans:1994cb}
  J.~M.~Evans,
{\sl Supersymmetry algebras and Lorentz invariance for d = 10 superYang-Mills},
  Phys.\ Lett.\  B {\bf 334}, 105 (1994)
  [arXiv:hep-th/9404190].
  %%CITATION = PHLTA,B334,105;%%

%\cite{Baulieu:2007ew}
\bibitem{Baulieu:2007ew}
  L.~Baulieu, N.~J.~Berkovits, G.~Bossard and A.~Martin,
{\sl Ten-dimensional super-Yang-Mills with nine off-shell supersymmetries},
  Phys.\ Lett.\  B {\bf 658}, 249 (2008)
  [arXiv:0705.2002 [hep-th]].
  %%CITATION = PHLTA,B658,249;%%

%\cite{Vassilevich:2003xt}
\bibitem{Vassilevich:2003xt}
  D.~V.~Vassilevich,
{\sl Heat kernel expansion: User's manual},
  Phys.\ Rept.\  {\bf 388}, 279 (2003)
  [arXiv:hep-th/0306138].
  %%CITATION = PRPLC,388,279;%%

%\cite{Voros:1986vw}
\bibitem{Voros:1986vw}
  A.~Voros,
{\sl Spectral functions, special functions and Selberg zeta function},
  Commun.\ Math.\ Phys.\  {\bf 110}, 439 (1987).
  %%CITATION = CMPHA,110,439;%%

\bibitem{Klebanov:1999rd}
  I.~R.~Klebanov and N.~A.~Nekrasov,
  {\sl Gravity duals of fractional branes and logarithmic RG flow},
  Nucl.\ Phys.\  B {\bf 574} (2000) 263
  [arXiv:hep-th/9911096].
  %%CITATION = NUPHA,B574,263;%%


\bibitem{ASduality}
  P.~C.~Argyres and N.~Seiberg,
{\sl S-duality in N=2 supersymmetric gauge theories},
  JHEP {\bf 0712} (2007) 088
  [arXiv:0711.0054 [hep-th]].
  %%CITATION = JHEPA,0712,088;%%

\bibitem{kazakov}
  D.~V.~Boulatov and V.~A.~Kazakov,
{\sl The Ising Model On Random Planar Lattice: The Structure Of Phase Transition
  And The Exact Critical Exponents},
  Phys.\ Lett.\  {\bf 186B} (1987) 379.
  %%CITATION = PHLTA,B186,379;%%

\bibitem{Gross}
D.J. Gross, I.R. Klebanov, A.V. Matytsin and A.V. Smilga, {\sl Screening vs. Confinement in 1+1 Dimensions}, [arXiv:hep-th/9511104].

\bibitem{screen2}
%\bibitem{Bardeen:1998eq}
  W.~A.~Bardeen, A.~Duncan, E.~Eichten and H.~Thacker,
{\sl Quenched approximation artifacts: A Study in two-dimensional QED},
  Phys.\ Rev.\  D {\bf 57} (1998) 3890.
  %%CITATION = PHRVA,D57,3890;%%

\bibitem{screen3}
%\bibitem{Trottier:1998nk}
  H.~D.~Trottier,
{\sl String breaking by dynamical fermions in lattice QCD: From three to  four
  dimensions},
  Phys.\ Rev.\  D {\bf 60} (1999) 034506
  [arXiv:hep-lat/9812021].
  %%CITATION = PHRVA,D60,034506;%%

\bibitem{screen4}
%\bibitem{Duncan:2000kr}
  A.~Duncan, E.~Eichten and H.~Thacker,
{\sl String breaking in four dimensional lattice QCD},
  Phys.\ Rev.\  D {\bf 63} (2001) 111501
  [arXiv:hep-lat/0011076]:\\
  %%CITATION = PHRVA,D63,111501;%%
%\bibitem{Bernard:2001tz}
  C.~W.~Bernard {\it et al.},
  {\sl Zero temperature string breaking in lattice quantum chromodynamics},
  Phys.\ Rev.\  D {\bf 64} (2001) 074509
  [arXiv:hep-lat/0103012]:\\
  %%CITATION = PHRVA,D64,074509;%%
%\bibitem{Bali:2005fu}
  G.~S.~Bali, H.~Neff, T.~Duessel, T.~Lippert and K.~Schilling  [SESAM
                  Collaboration],
  {\sl Observation of string breaking in QCD},
  Phys.\ Rev.\  D {\bf 71} (2005) 114513
  [arXiv:hep-lat/0505012].
  %%CITATION = PHRVA,D71,114513;%%

\bibitem{affleck}
For a modern review, see I. Affleck, {\sl Conformal field theory approach to Kondo effect},
[arXiv:cond-mat/9512099] and original references therein.

\bibitem{giveonkutasov}
A.~Giveon and D.~Kutasov,
  {\sl Brane dynamics and gauge theory},
  Rev.\ Mod.\ Phys.\  {\bf 71} (1999) 983
  [arXiv:hep-th/9802067].
  %%CITATION = RMPHA,71,983;%%

\bibitem{lineardilaton}
  S.~J.~Rey,
  {\sl The confining phase of superstrings and axionic strings},
  Phys.\ Rev.\  D {\bf 43} (1991) 526;\\
  %%CITATION = PHRVA,D43,526;%%
  C.~G.~.~Callan, J.~A.~Harvey and A.~Strominger,
  {\sl Worldbrane actions for string solitons},
  Nucl.\ Phys.\  B {\bf 367} (1991) 60.
  %%CITATION = NUPHA,B367,60;%%

 \bibitem{exactCFT}
  S.~J.~Rey,
 {\sl Axionic string instantons and their low-energy implications},
 in 'Workshop on Superstring and Particle Theory' eds. L. Clavelli and B. Harms, pp. 291-300
 (1989, World Scientific, Singapore);\\
  %%CITATION = C89/11/08;%%
C.~G.~.~Callan, J.~A.~Harvey and A.~Strominger,
  {\sl World sheet approach to heterotic instantons and solitons},
  Nucl.\ Phys.\  B {\bf 359} (1991) 611;\\
  %%CITATION = NUPHA,B359,611;%%
S.~J.~Rey,
  {\sl On string theory axionic strings and instantons},
in 'APS-DPF Annual Meeting' eds. D. Axen, D. Bryman and M. Comyn,
pp. 876-881 (1991, World Scientific, Singapore);\\
%%CITATION = C91-08-18;%%
C.~G.~.~Callan, J.~A.~Harvey and A.~Strominger,
  {\sl Supersymmetric string solitons}.
  arXiv:hep-th/9112030.
  %%CITATION = HEP-TH/9112030;%%

\bibitem{kutasov}
  S.~Elitzur, A.~Giveon, D.~Kutasov, E.~Rabinovici and G.~Sarkissian,
{\sl D-branes in the background of NS fivebranes},
  JHEP {\bf 0008} (2000) 046
  [arXiv:hep-th/0005052].
  %%CITATION = JHEPA,0008,046;%%

\bibitem{GK2}
  A.~Giveon and D.~Kutasov,
  {\sl Little string theory in a double scaling limit},
  JHEP {\bf 9910} (1999) 034
  [arXiv:hep-th/9909110];\\
  %%CITATION = JHEPA,9910,034;%%
 A.~Giveon and D.~Kutasov,
 {\sl Comments on double scaled little string theory},
  JHEP {\bf 0001} (2000) 023
  [arXiv:hep-th/9911039].
  %%CITATION = JHEPA,0001,023;%%

\bibitem{gaiotto}
  D.~Gaiotto,
  {\sl N=2 dualities},
  arXiv:0904.2715 [hep-th].
  %%CITATION = ARXIV:0904.2715;%%

%\cite{Gaiotto:2009gz}
\bibitem{gaiottomaldacena}
  D.~Gaiotto and J.~Maldacena,
  {\sl The gravity duals of N=2 superconformal field theories},
  arXiv:0904.4466 [hep-th].
  %%CITATION = ARXIV:0904.4466;%%

%\cite{Kawai:2007ek}
\bibitem{Kawai:2007ek}
  H.~Kawai and T.~Suyama,
{\sl AdS/CFT Correspondence as a Consequence of Scale Invariance},
  Nucl.\ Phys.\  B {\bf 789}, 209 (2008)
  [arXiv:0706.1163 [hep-th]].
  %%CITATION = NUPHA,B789,209;%%

%\cite{Kawai:2007eg}
\bibitem{Kawai:2007eg}
  H.~Kawai and T.~Suyama,
{\sl Some Implications of Perturbative Approach to AdS/CFT Correspondence},
  Nucl.\ Phys.\  B {\bf 794}, 1 (2008)
  [arXiv:0708.2463 [hep-th]].
  %%CITATION = NUPHA,B794,1;%%

%\cite{Berkovits:2007rj}
\bibitem{Berkovits:2007rj}
  N.~Berkovits and C.~Vafa,
{\sl Towards a Worldsheet Derivation of the Maldacena Conjecture},
  JHEP {\bf 0803}, 031 (2008)
  [AIP Conf.\ Proc.\  {\bf 1031}, 21 (2008)]
  [arXiv:0711.1799 [hep-th]].
  %%CITATION = APCPC,1031,21;%%

%\cite{Azeyanagi:2008mi}
\bibitem{Azeyanagi:2008mi}
  T.~Azeyanagi, M.~Hanada, H.~Kawai and Y.~Matsuo,
  {\sl Worldsheet Analysis of Gauge/Gravity Dualities},
  arXiv:0812.1453 [hep-th].
  %%CITATION = ARXIV:0812.1453;%%

%\cite{Kapustin:2009kz}
\bibitem{ABJMlocalization}
  A.~Kapustin, B.~Willett and I.~Yaakov,
  {\sl Exact Results for Wilson Loops in Superconformal Chern-Simons Theories with
  Matter},
  arXiv:0909.4559 [hep-th];\\
  %%CITATION = ARXIV:0909.4559;%%
  T.~Suyama,
{\sl On Large N Solution of ABJM Theory},
  arXiv:0912.1084 [hep-th];\\
  %%CITATION = ARXIV:0912.1084;%%
  N.~Drukker and D.~Trancanelli,
{\sl A supermatrix model for N=6 super Chern-Simons-matter theory},
  arXiv:0912.3006 [hep-th];\\
  %%CITATION = ARXIV:0912.3006;%%
  M.~Marino and P.~Putrov,
{\sl Exact Results in ABJM Theory from Topological Strings},
  arXiv:0912.3074 [hep-th].
  %%CITATION = ARXIV:0912.3074;%%

\bibitem{israel}
 D.~Israel,
 {\sl Non-critical string duals of N = 1 quiver theories},
  JHEP {\bf 0604} (2006) 029
  [arXiv:hep-th/0512166].
  %%CITATION = JHEPA,0604,029;%%



\end{thebibliography}
\end{document}